\documentclass[11pt]{article}
\usepackage{textcomp}
\usepackage{pdfsync}
\usepackage{fancyhdr}
\usepackage{amssymb}
\usepackage{amsmath}

\usepackage{graphicx}
\usepackage{latexsym}
\usepackage{appendix}
\usepackage{srcltx}
\def\Tr{{\rm Tr}}

\def\CD{{\cal D}}
\def\CE{{\cal E}}

\def\CK{{\cal K}}

\def\CO{{\cal O}}

\def\CT{{\cal T}}

\def\centeron#1#2{{\setbox0=\hbox{#1}\setbox1=\hbox{#2}\ifdim
   \wd1>\wd0\kern.48\wd1\kern-.48\wd0\fi
   \copy0\kern-.48\wd0\kern-.48\wd1\copy1\ifdim\wd0>\wd1
   \kern.48\wd0\kern-.48\wd1\fi}}

\def\CMP{Commun. Math. Phys.~}
\def\JHEP{JHEP~}

\def\PRL{Phys. Rev. Lett.~}
\def\PR {Phys. Rev.~}
\def\CQG {Class. Quant. Grav.~}
\def\PL {Phys. Lett.~}
\def\NP {Nucl. Phys.~}

\newcommand{\beq}{\begin{equation}}
\newcommand{\eeq}{\end{equation}}
\newcommand{\bea}{\begin{eqnarray}}
\newcommand{\eea}{\end{eqnarray}}
\newcommand{\ba}{\begin{array}}
\newcommand{\ea}{\end{array}}

\newcommand{\p}{\partial}
\newcommand{\nn}{\nonumber}
\newcommand{\bp}{\bar{\partial}}

\newcommand{\half}{\frac{1}{2}}

\newcommand{\bz}{{\bar{z}}}

\begin{document}

\hskip3cm

 \hskip12cm{CQUeST-2011-0451}
\vskip3cm

\begin{center}
 \LARGE \bf  Holographic Renormalization and      Stress Tensors\\
in  New Massive Gravity
\end{center}

\vskip2cm

\centerline{\Large \Large Yongjoon
Kwon$^{1}$  \,,~~Soonkeon
Nam$^{2}$\,, ~~Jong-Dae
Park$^{3}$\,, ~~Sang-Heon
Yi$^{4}$ }

\hskip2cm

\begin{quote}
Department of Physics and Research Institute of Basic Science, Kyung
Hee University, Seoul 130-701, Korea$^{1,2,3}$

Center for Quantum Spacetime, Sogang University, Seoul 121-741,
Korea$^4$
\end{quote}

\hskip2cm

\vskip2cm

\centerline{\bf Abstract} We obtain  holographically renormalized
boundary stress tensor with the emphasis on a special point in the
parameter space  of  three dimensional new massive gravity, using
the  so-called Fefferman-Graham coordinates with relevant counter
terms.   Through the linearized equations of motion with a standard
prescription, we also obtain correlators among these stress tensors.
We argue that the self-consistency of holographic renormalization
determines counter terms up to unphysical ambiguities. Using this
renormalized stress tensor in  Fefferman-Graham coordinates, we
obtain the central charges of dual CFT, and   mass and angular
momentum of some $AdS$ black hole solutions.  These results are
consistent with the previous ones obtained by other methods.   In
this study on the Fefferman-Graham expansion of new massive gravity,
some aspects of higher curvature gravity are revealed.\\
\underline{\hskip12cm}\\
${}^{1}$emwave@khu.ac.kr~~  ${}^{2}$nam@khu.ac.kr~~  ${}^{3}$jdpark@khu.ac.kr ~~ ${}^{4}$shyi@sogang.ac.kr

\thispagestyle{empty}
\renewcommand{\thefootnote}{\arabic{footnote}}
\setcounter{footnote}{0}

\newpage

\section{Introduction}
It is natural to consider higher curvature corrections in gravity
theories beyond an Einstein-Hilbert  term with(out) a cosmological
constant    in the viewpoint of string theory or quantum gravity,
though there are some attempts, for instance, loop quantum gravity
approach, to make pure Einstein gravity into self-consistent quantum
theory without such corrections. However, it is not  easy task to
perform a detailed analysis of higher curvature gravity in  more
than four dimensions, partly because of  the complicated dynamics
and the difficulty in obtaining analytic solutions. Moreover, the
higher derivative terms with general covariance may conflict with
unitarity in many cases.   The situation is, more or less, different
in lower dimensional higher curvature gravity theories since the
reduction in degrees of freedom simplifies the dynamics and the
existence of analytic black hole solutions leads to some analytic
results.   In particular,  the three dimensional higher curvature
theory has non-empty contents and it may admit non-trivial black
hole solutions of $AdS$  asymptotics. Therefore, it is useful route
to study three dimensional  gravity theory to obtain insights on
gravity with higher derivative terms in conjunction with $AdS/CFT$
correspondence~\cite{Maldacena:1997re, Witten:1998qj}.   It is also
notable  that the reliable quantum computation in gravity without
supersymmetry is performed only in three
dimensions~\cite{Witten:1988hc} and that
Kerr/CFT~\cite{Guica:2008mu} is inspired by  $AdS/CFT$
correspondence on three dimensions.

In last few years, the specific higher curvature gravity theory in
three dimensions, now known as new massive gravity (NMG)
\cite{Bergshoeff:2009hq, Bergshoeff:2009aq,  Townsend:2009,
Bergshoeff:2009fj}, has drawn some interests and leads to another
realization of  the $AdS/CFT$ correspondence.  The original
motivation of NMG is the non-linear completion of the Fierz-Pauli
massive graviton theory, which is not yet done in higher dimensions.
Concretely, the  simplest form of NMG is composed of a curvature
scalar term with(out) a cosmological constant and the other  term
which is the specific combination of curvature scalar square and
Ricci tensor square.  This parity-even NMG shares some aspects with
the so-called topologically massive gravity (TMG) \cite{Deser:1982,
Deser:1981wh}  in three dimensions, which is composed of a curvature
scalar term with a cosmological constant   and a gravitational
Chern-Simons term. One of common aspects in these two theories is
the existence of $AdS$ `vacuum'
 solution.  Soon after the introduction of NMG, it was
realized that $AdS$ space may be allowed as a solution, and so some
studies \cite{Clement:2009gq, Oliva:2009ip, Giribet:2009qz,
Ghodsi:2010, Liu:2009kc,Ahmedov:2010em} are done along the route of
$AdS/CFT$ correspondence in this new setup, similarly to  the case
in TMG \cite{Nutku:1993, Strominger:2008,  Clement:2008,
Anninos:2008fx}. Basically, all these studies do not require the
supersymmetry as a crucial ingredients and so may be regarded as a
test or the realization of $AdS/CFT$ correspondence just with
bosonic degrees of freedom.

Aside from some similarity to TMG, NMG has several different aspects
from TMG.  Because of these,  various methods developed for TMG may
not be applied directly to NMG.   One peculiar thing in NMG is that
the radius of the $AdS$ space is not identical with a cosmological
constant, since the higher curvature term may also play the role of
an effective cosmological constant.  Another notable difference
resides in the fact that there is no guarantee of  the completeness
of NMG from the very beginning. On the contrary,  there are some
suggestions of the completeness of TMG  at a certain special point
in the parameter space~\cite{Strominger:2008}, though there are
several arguments against these~\cite{Grumiller:2008qz}. It turns
out that this incompleteness in NMG is not just drawbacks  but
allows  the extension of NMG to even higher curvature gravity
theories \cite{Sinha:2010,Gullu:2010}. The underlying principle in
this extension is again the renowned $AdS/CFT$ correspondence. The
important observation in this development is the compliance of the
NMG with the holographic c-theorem \cite{Sinha:2010, Myers:2010tj},
which is a specific incarnation of $AdS/CFT$ correspondence in this
context. The consistency with the holographic c-theorem might be  an
important ingredient for quantum gravity if we suppose that
$AdS/CFT$ correspondence has a deep meaning in quantum gravity.
Because the simplest form of NMG is uniquely determined by the
holographic c-theorem, it is a proper step  to study various aspects
of NMG in this simplest case. In this paper, we will focus on the
simplest form of NMG, and will call this form just as NMG in the
following.

Though a primitive form of $AdS/CFT$ correspondence was found in
three dimensions by Brown and Henneaux \cite{Brown:1986nw}, the
concrete version of $AdS/CFT$ correspondence from string theory
was realized in Maldacena's work \cite{Maldacena:1997re} and
afterwards the prescription for the matching between partition
functions are established \cite{Witten:1998qj}. Specifically,
on-shell bulk gravity action on $AdS$ space becomes a functional of
boundary values  after solving  equations of motion (EOM) of the
bulk modes in terms of the boundary values. Then,    the boundary
values of non-normalizable modes of bulk gravity theory  play  the
role of sources of corresponding operators in  the dual CFT side,
and one can evaluate  (at least,  large  't Hooft coupling)
correlators of the corresponding   operators through gravity
computations.  In particular, the boundary stress tensor in CFT
couple to the boundary metric which is naturally identified with the
induced metric from the bulk one. Therefore,  those can be obtained
just from the pure gravity modes.  As usual in other operators in
field theories,  stress tensor requires some renormalization  for
finite results. This process is also realized in the side of bulk
gravity through $AdS/CFT$ correspondence, which is now coined as
{\it holographic renormalization} \cite{Henningson:1998gx,
Hyun:1998vg, Balasubramanian:1999re, Solodukhin:2001,
Skenderis:2002wp, Kraus:2008}.

In a gravity theory with a boundary, one may need a boundary term in
the action to ensure the well-defined variational principle.  In the
case of pure Einstein  gravity,  this boundary term  has been called
the Gibbons-Hawking (GH) term~\cite{Gibbons:1976ue}, which turns out
to be just the extrinsic curvature of the boundary.   With the
equipment of this GH term    boundary stress tensor  was introduced
a long time ago by Brown and York~\cite{Brown:1992br}, which
contains divergence in the case of  $AdS$ space.  According to
$AdS/CFT$ dictionary,  this fact has been interpreted as the counter
part of the necessity of the renormalization in the boundary CFT,
and so some counter terms, which are local in the boundary and
consistent with the variational principle, are constructed and lead
to finite results for boundary stress tensor.  This holographic
renormalization is one of the highly successful realization of
$AdS/CFT$ correspondence. Though this construction is conceptually
transparent,  it has been not yet implemented for most of   higher
curvature gravity theories since there are  several challenging
steps to construct holographically renormalized boundary stress
tensor. For instance, a relevant GH term  has not yet been found for
most of higher derivative gravity theories.  However, one may note
that  holographic renormalization  was implemented  successfully in
TMG without information about GH term~\cite{Skenderis:2009nt}.

In NMG the generalized GH term was obtained with the aim of
$AdS/CFT$ correspondence~\cite{Hohm:2010jc} and holographically
renormalized stress tensor was also studied for   mass and angular
momentum of some $AdS$ black holes \cite{Giribet:2010ed}.
Furthermore, some correlators of renormalized stress tensor are
obtained at the so-called {\it critical point}, revealing
characteristic of logarithmic CFT (LCFT) from bulk $\log$
modes~\cite{Alishahiha:2010bw, Grumiller:2009sn, Grumiller:2010tj}.
Even with these studies  there are some missing parts in the
analysis and it is also desirable to study holographic
renormalization in the unified manner.  One may note that  there is
another special point other than critical point in NMG, which is
related to the existence of   the so-called {\it new type black
holes} \cite{Bergshoeff:2009aq, Oliva:2009ip, Nam:2010ma,
Maeda:2010, Kwon:2011ey}. At this special  point, it has been known
that a different fall-off of bulk modes is allowed. While  mass and
angular momentum of new type black holes are identified as conserved
charges \cite{Nam:2010ma} and consistent with  renormalized stress
tensor~\cite{Giribet:2010ed}, the central charge of dual CFT in this
case has been obtained indirectly through Cardy
formula~\cite{Oliva:2009ip} or central charge function
formalism~\cite{Nam:2010dd}.  According to holographic
renormalization or $AdS/CFT$ correspondence, all these quantities
should be obtained uniformly from renormalized boundary stress
tensor.

In this paper, we study the holographic renormalization of boundary
stress tensor in the so-called   Fefferman-Graham coordinates
\cite{FG} which are especially suitable to the purpose and relevant
to obtaining several physical quantities in the unified manner.  We
argue that the consistency of holographic renormalization determines
the relevant counter terms up to unphysical ambiguities. This is one
of
  improvements over the previous related studies. The
organization of the paper is as follows. In the next section
Fefferman-Graham coordinates are adopted and Brown-York boundary
tensor is obtained. In section 3 several possible counter terms are
considered and the corresponding renormalized boundary stress tensor
is obtained.  Accordingly,  the central charge of dual CFT is
obtained as a trace anomaly of renormalized stress tensor. Using
linearized EOMs in NMG  with  the standard prescription in the
$AdS/CFT$ correspondence, we obtain the correlators of boundary
stress tensor in section 4, which give us the central charge of dual
CFT and some other information. In this section we argue that the
consistency of holographic renormalization determines counter terms
up to unphysical ambiguities.   In section 5, through renormalized
stress tensor dictated  by  holographic renormalization  we   obtain
mass and angular momentum of some $AdS$ black holes, which  are
consistent with previous results obtained by other methods.   In the
final section we summarize our results with some comments. In
appendixes, useful formulae are collected, which are used in the
main text.


\section{Fefferman-Graham Expansion in NMG}
After introduced as  a completion of Fierz-Pauli linear graviton
theory, NMG was recognized as a consistent form  with holographic
c-theorem and was in fact uniquely determined by it. This uniqueness
of Lagrangian provides a strong  support for $AdS/CFT$
correspondence of any kind of $AdS$ solutions in this theory.
Therefore, it is desirable to analyze the boundary stress tensor for
$AdS$ asymptotics in the unified manner for any  parameter values in
the Lagrangian. For this purpose  it turns out that the so-called
Fefferman-Graham coordinates are appropriate ones. The Lagrangian of
NMG we will consider in the following is given by
\begin{equation}\label{NMG}
S =\frac{\xi}{2\kappa^2}\int d^3x\sqrt{-g}\bigg[ \sigma R +
\frac{2}{\ell^2} + \frac{1}{m^2}\CK  \bigg]\,,
\end{equation}
where $\xi$ and $\sigma$ take $1$ or $-1$,  $2\kappa^2=16\pi G$, and
$\CK$ is a specific combination of scalar curvature square  and
Ricci tensor square defined by \beq
 \CK = R_{\mu\nu}R^{\mu\nu} -\frac{3}{8}R^2\,.
\eeq
Our convention is such that $m^2$ is always positive but the
cosmological constant $\ell^2$ has no such restriction\footnote{We
have introduced  $\xi$ for  the various sign choice of  terms in the
action. but it will be set unity in the following.}. The equations
of motion  of NMG are given by
\begin{equation}\label{eom}
\CE_{\mu\nu} =\xi \Big[
 \sigma G_{\mu\nu} - \frac{1}{\ell^2}g_{\mu\nu} + \frac{1}{2m^2}\CK_{\mu\nu}\Big]
        =0\,,
\end{equation}
 where
\begin{equation}
 \CK_{\mu\nu} = g_{\mu\nu}\Big(3R_{\alpha\beta}R^{\alpha\beta}-\frac{13}{8}R^2\Big)
                + \frac{9}{2}RR_{\mu\nu} -8R_{\mu\alpha}R^{\alpha}_{\nu}
                + \half\Big(4\CD^2R_{\mu\nu}-\CD_{\mu}\CD_{\nu}R
                -g_{\mu\nu}\CD^2R\Big)\,. \label{Ktensor}
\end{equation}

In the above Lagrangian of NMG there is an additional parameter
$m^2$ of mass dimension two in front of $\CK$ term along with the
gravitational constant  $G$ and cosmological constant $\ell$. In NMG
$AdS$ space is allowed as a solution and its radius $L$, which is
always positive, is related to parameters of the  Lagrangian as
\beq \label{rel}
  \frac{1}{L^2} = 2m^2 \left[\sigma \pm \sqrt{1-\frac{1}{m^2 \ell^2} }\right]\,.
\eeq
Instead of a cosmological constant  $\ell$, we will use $L$ as a
basic parameter in the following since we will confine ourselves to
the case of asymptotically $AdS$ space. Note that the dimensionless
combinations of three parameters can be represented by $L/G$ and
$m^2L^2$ which are positive quantities.

Because of the additional parameter $m^2$,  there are  several
special points in the parameter space, one of which is now coined as
{\it critical point}~\cite{Lu:2011zk,Deser:2011xc,Myung:2011uy}. At
this point which is given by the parameter condition $-\sigma =
2m^2L^2=1 $ $(m^2 \ell^2=-1/3)$ in our convention, it was shown that
there may be $\log$ modes in the bulk gravity which corresponds to
the non-unitary LCFT at  the boundary. These $\log$ tails of bulk
modes differ from Brown-Henneaux fall-off boundary condition and
therefore the correspondence at this point is regarded as the
extension of the standard $AdS/CFT$ correspondence. Another
interesting point of parameter space is given by $\sigma= 2m^2L^2=1$
$(m^2 \ell^2=1)$. At this special point, there are $AdS$ black hole
solutions which have different fall-off behaviors from
Brown-Teitelboim-Zanelli (BTZ) ones \cite{btz}. As in the case of
the above critical point, it is reasonable to perform further
analysis at this special point in the viewpoint of holographic
renormalization, which is the main focus in this paper.

Through the $AdS/CFT$ correspondence, holographically renormalized
boundary stress tensor corresponds  to one point function of stress
tensor in the dual CFT, and so can be used to obtain the central
charge of the dual CFT. It has been known that those can also be
used to obtain  mass and angular momentum of relevant $AdS$ black
holes. The holographically renormalized stress tensor may be
introduced in the following way. First, let us obtain the Brown-York
boundary stress tensor for which one needs the GH boundary term for
the well-defined variational procedure. This Brown-York stress
tensor is usually divergent because of the nature of asymptotic
$AdS$ space. And then, by a suitable choice of local counter terms
in the boundary, one can introduce the renormalized stress tensor.
To apply this procedure   with $AdS$ asymptotics, it has been  known
that the Fefferman-Graham coordinates are especially appropriate.

Now, let us briefly explain the  Fefferman-Graham expansion  in our
convention. Asymptotically $AdS$ space may be put in the following
metric form which is useful for our purpose
\bea ds^2 &=& L^2 \gamma_{\mu \nu}dx^{\mu} dx^{\nu} =L^2\Big[ d\eta^2 + \gamma_{ij}dx^idx^j\Big]\,,  \nn \\  \label{FGexp}  \\
          \gamma_{ij} &=& e^{2\eta} g^{(0)}_{ij} + e^{\eta}  g^{(1)}_{ij}  +  g^{(2)}_{ij} +  e^{-\eta} g^{(3)}_{ij} +   e^{-2\eta}g^{(4)}_{ij} + \CO(e^{-3\eta})\,.  \nn\eea
Non-vanishing Christoffel symbols in the above Fefferman-Graham
coordinates are given by
\[ \Gamma^{\eta}_{ij} = -K_{ij}\,, \qquad \Gamma^{i}_{\eta j} = K^{i}_{j}\,, \qquad \Gamma^{k}_{ij} = {}^{(2)}\Gamma^{k}_{ij} (\gamma)\,, \]
where $K_{ij}$ is the extrinsic curvature tensor given by $K_{ij}
=\half \p_{\eta} \gamma_{ij}$ in this coordinate system and
${}^{(2)}\Gamma^{k}_{ij}$ is two dimensional  Christoffel symbols
given by the metric $\gamma$.   In the following   $g_{(0)}$ is
taken as the boundary metric for two dimensional CFT and geometrical
quantities at the boundary are denoted with the script $(0)$.  For
instance, the boundary scalar curvature is denoted as $R_{(0)}$.
 In these coordinates Ricci tensor  is given by
\bea R^{\eta}_{~ \eta} &=& - K' -K_{ij}K^{ij}\,,   \\
       R^{\eta}_{~ i} &=& \nabla^jK_{ji} -\nabla_iK\,, \nn \\
      R^{i}_{~ j} &=& {}^{(2)}R^{i}_{~j} -(K^{i}_{j})' -KK^{i}_{~j}\,, \nn
       \\  R &=& {}^{(2)}R-K_{ij}K^{ij} -K^2 -2K'\,, \nn
       \eea
where {}{}$'$  denotes the differentiation with respect to $\eta$,
$K\equiv \gamma^{ij}K_{ij}$, ${}^{(2)}R^{i}_{~j}$ is the two
dimensional Ricci tensor, and $\nabla$ denotes the covariant
derivative with respect to the two dimensional metric $\gamma$.

The above coordinates are known as Fefferman-Graham coordinates (or
expansion) and extremely useful for  holographic renormalization.
In the following,   the fluctuation modes as well as background
space are taken in these coordinates and  the $AdS$ radius is set as
the unity for the convenience, {\it i.e.} $L=1$ which can be
reinstated by a simple dimensional reasoning. One may notice that we
have included $g_{(k)}$ terms of odd $k$ deliberately
in~(\ref{FGexp}), which turns out to be related to higher curvature
effects. In the case of usual pure Einstein gravity and TMG,
$g_{(k)}$ terms of odd $k$ have been dropped because they vanish by
EOM.    Contrary to these, we will show that  this  is not the case
at the  special parameter point in NMG.  Odd $k$ terms are not taken
in previous analysis~\cite{Alishahiha:2010bw} in NMG since a
critical point not our special point is the main interest.   One may
note that the relevance of $g_{(k)}$ terms of odd $k$ is mentioned
in a certain matter-coupled three dimensional gravity without a
detailed analysis~\cite{Skenderis:2009nt,Berg:2001ty}.

For completeness, let us briefly review holographic renormalization
and renormalized stress tensor  in the case of pure
Einstein-Hilbert gravity with the focus on the central charge of
dual CFT. The boundary renormalized stress tensor, $T^{ij}$ is
introduced  as
\beq \delta (S_{Bulk} + S_{GH} + S_{c.t.}) = \half \int d^2x
\sqrt{-g_{(0)}}\, T^{ij}\delta g^{(0)}_{ij}\,,  \eeq
where $S_{Bulk}$ is a bulk gravity action, $S_{GH}$ is a GH boundary
term and $S_{c.t.}$ is the so-called counter term which renders
finite the boundary stress tensor. Two terms $S_{Bulk}$ and $S_{GH}$
give  the Brown-York boundary stress tensor and the counter term
does a counter boundary stress tensor. In the above Fefferman-Graham
coordinates one can see that the renormalized boundary stress tensor
can be obtained by the following formula
\beq 8\pi G\, \sqrt{-g_{(0)}}\,T^{ij} = 8\pi G
e^{2\eta}\sqrt{-\gamma}\Big[ T^{ij}_{BY} +
T^{ij}_{c.t.}\Big]_{\eta\rightarrow\infty}   \,,\eeq
 where $T^{ij}_{BY}$ is Brown-York boundary stress tensor and $T^{ij}_{c.t.}$ is counter one. One may note that the divergences of stress tensor without a counter term  come from  the combined quantity $e^{2\eta}\sqrt{-\gamma}\, T_{BY}$ in   our convention.   For brevity we call these divergences  as those of Brown-York tensor in the following.

In the pure Einstein gravity case almost unique choice of counter
term, which is local at the boundary, is just the boundary
cosmological constant with a suitable coefficient\footnote{It is
well known that there should be  another term of the form $(1/16\pi
G)\eta  \int d^2x \sqrt{-\gamma} R$, to render  on-shell action
finite. However, this term is topological in two dimensions and so
irrelevant to stress tensor computation in the following. See
Appendix D.}~\cite{Kraus:2008, Skenderis:1998}
\[\qquad S_{c.t.} = -\frac{1}{8\pi G}\int d^2x \sqrt{-\gamma}\,. \]
In this case Brown-York and counter stress tensor are given by
\[
8\pi G\, T^{ij}_{BY} = K\gamma^{ij} -K^{ij}\,, \qquad 8\pi G\,
T^{ij}_{c.t.} = -\gamma^{ij}\,.
\]
Then, the renormalized stress tensor is given explicitly by
\beq
 8\pi G\, \sqrt{-g_{(0)}}\,T^{ij} =   e^{2\eta}\sqrt{-\gamma}\Big[(K-1)\gamma^{ij} -K^{ij}\Big]_{\eta\rightarrow\infty}\,,     \eeq
which lead to
\[  8\pi G\, T^{ij}    =  g^{ij}_{(2)} -(\Tr\, g_{(2)})g^{ij}_{(0)} \,.     \]
In this expression all quantities are raised or lowered by the
boundary metric $g_{(0)}$ and $`\Tr$' denotes the contraction by
$g_{(0)}$. Note that $g_{(1)}=0$ and $R_{(0)}+2\Tr\,g_{(2)}=0$ from
the EOMs.  One can see that these give  us the correct mass and
angular momentum of BTZ black holes and the central charge of dual
CFT.  Concretely for the central charge,  one can see that the
contraction of renormalized stress tensor is given by
\cite{Kraus:2008, Skenderis:1998}
\[
8\pi G\, T = -\Tr\, g_{(2)}\,.
\]
Using EOM in Fefferman-Graham coordinates and recalling $\langle
T\rangle=  (c/24\pi )R_{(0)}$ in two dimensional conformal anomaly
through the identification of boundary CFT stress tensor with
holographic one,  one obtains  the central charge of dual CFT as
$c=3/2G$.

Now let us  recall  the generalized GH term in NMG. To obtain the
generalized GH term in NMG,   Hohm and Tonni has used the auxiliary
field approach\footnote{One may note  that our convention is
slightly different from~\cite{Hohm:2010jc}.}~\cite{Hohm:2010jc}.
In summary,   the above NMG action can be rewritten in terms of
auxiliary field $f_{\mu\nu}$ as
\beq S =\frac{1}{2\kappa^2}\int d^3x\sqrt{-g}\bigg[ \sigma R +
\frac{2}{\ell^2} +
f^{\mu\nu}G_{\mu\nu}-\frac{m^2}{4}\Big(f^{\mu\nu}f_{\mu\nu} -
f^2\Big)  \bigg]\,.  \label{Action} \eeq
In this representation the EOMs of this action are given by
\beq    \sigma G_{\mu\nu} -\frac{1}{\ell^2}g_{\mu\nu}  =
\CT^{B}_{\mu\nu}\,, \qquad   f_{\mu\nu}
=\frac{2}{m^2}\Big(R_{\mu\nu} - \frac{1}{4}R g_{\mu\nu}\Big)\,,
        \eeq
where \bea \!
     \CT^{B}_{\mu\nu} &=& \frac{m^2}{2}\Big[ f_{\mu\alpha} f^{\alpha}_{\nu} -f f_{\mu\nu} -\frac{1}{4}\Big(f^{\alpha\beta}f_{\alpha\beta} -f^2\Big)  g_{\mu\nu}\Big]  + \half f R_{\mu\nu} - \half Rf_{\mu\nu} -2 f_{\alpha (\mu}G^{\alpha}_{\nu)} + \half f^{\alpha\beta}G_{\alpha\beta} g_{\mu\nu} \nn \\
      && -\half \Big[ \CD^2f_{\mu\nu} + \CD_{\mu}\CD_{\nu} f -2\CD^{\alpha}\CD_{(\mu}f_{\nu)\alpha} + \Big(\CD_{\alpha}\CD_{\beta}f^{\alpha\beta} - \CD^2f\Big)g_{\mu\nu} \Big]\,.  \nn \eea
This form of NMG is useful to obtain the generalized Gibbons-Hawking
boundary term.  As the metric decomposition, auxiliary fields
$f^{\mu\nu}$ can be decomposed as
\bea\label{auxliary} f^{\mu\nu} = \left(\ba{cc} s & h^j \\ h^i &
f^{ij} \ea\right) \,.  \nn \eea
With this decomposition the generalized GH term was obtained   in
the form of
\beq
 S_{GH} = \frac{1}{2\kappa^2}\int d^2x \sqrt{-\gamma}\Big[2\sigma K + \hat{f}^{ij}K_{ij} - \hat{f}K\Big]\,,
\eeq
where the first term proportional to $\sigma$ is the GH term in pure
Einstein gravity case and $\hat{f}^{ij}$, $\hat{f}$ are given
in~\cite{Hohm:2010jc}.
With this result, Brown-York stress tensor for NMG is also obtained
in~\cite{Hohm:2010jc, Giribet:2010ed}. After some rearrangement
those can be written as
\beq \!
 8\pi G  T^{ij}_{BY} =   \Big(\sigma + \half \hat{s} -\half \hat{f}\Big)(K\gamma^{ij} - K^{ij})  - \nabla^{(i} \hat{h}^{j)}+ \half D_{\eta} \hat{f}^{ij}+ K^{(i}_{k} \hat{f}^{j) k}  +\gamma^{ij} \Big(\nabla_{k}\hat{h}^{k} -\half D_{\eta}\hat{f}\Big) \,,  \nn \eeq
where the first term proportional to $\sigma$ is just the Brown-York
tensor for the pure gravity. Note that  in the Fefferman-Graham
coordinate system $D_{\eta}$ becomes the ordinary derivative with
respect to $\eta$. That is to say, $D_{r}=\p_{\eta} $  and
~$\hat{}$~ has no effect in our case: $\hat{f} ={\gamma_{i j}
f^{ij}}$, $\hat{h}=h$, $\hat{s} =s$ and $f=\gamma_{\mu \nu}f^{\mu
\nu}={\hat f}+s$.   It is important to remind that $f^{i}_{j}$,
neither $f_{ij}$ nor $f^{ij}$, are taken as fundamental variables
to construct generalized GH term and accordingly Brown-York stress
tensor.

One can now see that Brown-York tensor,  $T^{ij}_{BY}$ expanded in
Fefferman-Graham coordinates is given by
\bea  \!
  8\pi G\,  T^{ij}_{BY} &=& e^{-2\eta} \Big(\sigma + \frac{1}{2m^2}\Big)  g^{ij}_{(0)}  \  -\half\, e^{-3\eta}\bigg[ \Big(\sigma + \frac{1}{2m^2}\Big)(\Tr\, g_{(1)})\, g^{ij}_{(0)}  + \Big(\sigma +\frac{3}{2m^2}\Big)\, g^{ij}_{(1)} \bigg] \nn\\
                     &&\!\!\!\!\!     \!\!\!\!\!       - e^{-4\eta}\bigg[\Big\{\frac{1}{4m^2}R_{(0)}  + \Big(\sigma + \frac{1}{m^2} \Big) \Tr\, g_{(2)} -\half \Big(\sigma + \frac{5}{8m^2}\Big)\Tr\, g^2_{(1)} - \frac{1}{16m^2} (\Tr\, g_{(1)})^2\Big\}\, g^{ij}_{(0)} \nn \\
                          &&~~~  -\frac{1}{2m^2} g^{ik}_{(1)}g^{j}_{(1)\, k} -\half\Big(\sigma + \frac{1}{m^2}\Big)(\Tr\, g_{(1)})\, g^{ij}_{(1)}   \bigg] +\CO(e^{-5\eta})\,. \label{BYTensors}
\eea
Note that there are two divergences  which come from the first and
the second terms in the above Brown-York tensor. Though there are
some proposals which  counter term is appropriate for the
renormalized boundary stress tensor in the case of NMG,  which will
be reviewed in the next section, it is desirable to investigate
counter terms and renormalized stress tensor in the viewpoint of
holographic renormalization with Fefferman-Graham coordinates.  It
turns out that  some features are transparent in  Fefferman-Graham
coordinates.

\section{Counter Terms and Renormalized Stress Tensors}
In this section we construct relevant counter terms in NMG and then
obtain renormalized stress tensor under the perspective of
holographic renormalization. This stress tensor gives us the correct
central charge of dual CFT and mass and angular momentum of $AdS$
black holes, which are consistent with  previous results obtained by
other methods.  There are some studies about holographically
renormalized stress tensor in some parameter regions, which are
briefly reviewed in the following.  Though renormalized stress
tensor at the special point of $\sigma=2m^2=1$ was
studied~\cite{Giribet:2010ed}, there are some gaps in the logic and
not sufficiently generic. For instance,  the central charge of dual
CFT is dealt with differently from mass and angular momentum of
$AdS$ black holes.  As mentioned in the introduction, all these
quantities can be  managed at one stroke in Fefferman-Graham
coordinates. Furthermore, this approach does not resort on a
particular black hole solutions rather uses just the generic
expansion form.    In addition to this improvement, Fefferman-Graham
expansion is very useful to obtain correlators of stress tensor
through holographic renormalization, which are presented in the next
section.

There have been two suggestions for relevant counter terms in
NMG~\cite{Hohm:2010jc, Giribet:2010ed}.  In our perspective,  the
first case corresponds to $g_{(1)} =0$ and the second one does to
$g_{(1)}\neq0$. Now, let us explain these choices of counter terms
in the viewpoint of our approach. The first prescription for counter
term is given in~\cite{Hohm:2010jc} for the so-called Brown-Henneaux
fall-off boundary conditions. The relevant counter term is taken  by
the boundary cosmological constant just like the pure gravity case
with a suitable coefficient
\[
S_{c.t.} =  -\Big(\sigma + \frac{1}{2m^2}\Big)  \frac{1}{8\pi G}
\int d^2x \sqrt{-\gamma} \,. \]
In this choice  of a counter term,    the leading divergence of the
Brown-York tensor can be canceled, which is the unique divergent
term under the condition of $g_{(1)}=0$. It was also shown that
renormalized stress tensor reproduces the correct central charge of
dual CFT using the asymptotic symmetry algebra, which is consistent
with the results by central charge function formula or by Cardy
formula. However, it is insufficient for the $g_{(1)} \neq 0$ case
as one can see the expression of Brown-York tensor in
Eq.~(\ref{BYTensors}). There is another divergence coming from the
$g_{(1)}$ part. This may be paraphrased as this counter term is
appropriate to Brown-Henneaux boundary conditions but it is
insufficient for weaker boundary conditions than Brown-Henneaux.

A different counter term for weaker fall-off boundary condition than
Brown-Henneaux one is taken in~\cite{Giribet:2010ed} as
\[   S_{c.t.} =  \frac{m^2}{2} \Big(\sigma + \frac{1}{2m^2}\Big)  \frac{1}{8\pi G} \int d^2x \sqrt{-\gamma}~ \hat{f}\,. \]
%
 At first sight, one may suspect that the chosen counter term  cancels only the leading divergence  and a part of the next leading one of Brown-York tensor, not the whole  of the next leading one.  However, one can see that the remnant potential divergence cancels completely,  either since    $g_{(1)}$  term vanishes by EOM at the generic parameter point  or since  the coefficient of next leading divergence vanishes  at the special point  of $\sigma=2m^2=1$.   Though the above choice of counter term cancels all the divergences in NMG at the special parameter point, its inference is neither  completely logical nor unique  since it depends on specific black hole solutions called new type black holes. Moreover, it was argued that one needs to consider rotating new type black holes for the complete determination.

%
%

  Meanwhile,  in holographic renormalization  black hole solutions are not essential to obtain central charges, as can be seen from the fact that  the appearance of $\log$ modes at the critical point cannot be deduced from known black hole solutions~\cite{Grumiller:2009sn, Grumiller:2010tj}. To overcome this logical drawback and clarify the meaning of freedom in counter term choice, let us consider generic counter terms  without resorting to  specific black hole solutions. Then,   it is legitimate to  consider the following generic counter terms
\[  S_{c.t.} =     \frac{1}{8\pi G} \int d^2x \sqrt{-\gamma}~(A +B \hat{f} +C\hat{f}^2 +Df_{ij}f^{ij})\,, \]
which lead to
\[
8\pi G\, T^{ij}_{c.t.} =   (A +B \hat{f}
+C\hat{f}^2+Df_{kl}f^{kl})\, \gamma^{ij} \,. \]
To obtain this expression, it is crucial to  recall that the
fundamental fields under the variation are $f^{\mu}_{\nu}$ neither
$f^{\mu\nu}$ nor $f_{\mu\nu}$, as advertized in~\cite{Hohm:2010jc} .

To cancel the divergence, one should take $A,B,C,D$  as
\bea \label{cds}
0&=& A -\frac{2}{m^2}B + \frac{4}{m^4}C + \frac{2}{m^4}D + \sigma + \frac{1}{2m^2} \,,   \\
\label{cds1}
        0 &=& \frac{1}{m^2}B -\frac{4}{m^4}C -\frac{2}{m^4}D - \frac{1}{2}\Big(\sigma + \frac{1}{2m^2}\Big)\,.   \eea
The first condition is needed for the cancelation of the leading
divergence in  Brown-York tensor, and the second one for the
cancelation of the next leading one. One may note that the second
condition is not necessary for the case of $g_{(1)}=0$, which is
implied by EOMs at the generic point in the parameter space. In
other words, there are ambiguities in choosing   counter terms at
the generic point in the parameter space. One can see that any
choice of $A,B,C,D$ satisfying the first condition gives us the same
results for on-shell boundary stress tensor.  To see this, note that
the renormalized boundary stress tensor  is  given by
\bea
 8\pi G\, T^{ij}  &=&    \Big(\sigma + \frac{1}{2m^2}\Big)g^{~~ ij}_{(2)} -  \bigg[ \frac{1}{4m^2}R_{(0)} + \Big(\sigma + \frac{1}{m^2}\Big) \Tr\, g_{(2)}  \nn \\
 &&~~~ \qquad \qquad \qquad~~~ - \Big(\frac{1}{m^2}B-\frac{4}{m^4}C-\frac{2}{m^4}D\Big)\Big(R_{(0)} +2\Tr\, g_{(2)}\Big)\bigg] g^{ ij}_{(0)}  \,. \label{ST1} ~~~~~
\eea
%
%
To impose on-shell condition, let us note that $\eta\eta$-components
of EOMs or the contracted one of EOMs up to relevant orders lead to
\beq R_{(0)}+2\Tr\, g_{(2)}=0\,. \eeq
Since EOMs in   NMG   are complicated by higher curvature terms,
some details about EOMs are relegated to appendixes. See there for
the derivation of the above. As a result, one obtains
\beq 8\pi G\, T^{ij} =  \Big(\sigma + \frac{1}{2m^2}\Big)\Big[g^{~~
ij}_{(2)} -(\Tr\, g_{(2)})g^{~~ ij}_{(0)}\Big]\,, \eeq
which shows us ambiguities in counter terms in this case.  This
phenomenon is a reminiscent of the renormalization scheme
independence in dual field theory.  The above renormalized stress
tensor  under the contraction  with the boundary metric $g_{(0)}$
leads to
\beq
  8\pi G\, T  = \frac{1}{2} \Big(\sigma + \frac{1}{2m^2}\Big) R_{(0)}\,,
\eeq
and  then central charge is obtained as
\beq  c =  \Big(\sigma + \frac{1}{2m^2}\Big) \frac{3}{2G}\,, \eeq
which reproduces the results in~\cite{Hohm:2010jc, Liu:2009bk}. This
shows us that the boundary cosmological constant as a counter term
is not unique choice but may be regarded as a minimal one.

In the case of $g_{(1)} \neq 0$, which is allowed only at the point
$\sigma = 2m^2 =1$ for the finiteness of the stress tensor
(\ref{rbst}), we should impose the second condition.  Even in this
case, there is still ambiguity in choosing the counter terms as can
be seen explicitly as follows. Firstly, under the above two
conditions
 the renormalized boundary stress tensor is given by
\bea \label{rbst} 8\pi G\, T^{ij} & =&
\frac{1}{2}e^{\eta}\Big(\sigma-\frac{1}{2m^2}\Big)g^{ij}_{(1)} +
\Big(\sigma + \frac{1}{2m^2}\Big)g^{ij}_{(2)} -\sigma g^{~~i}_{(1)\,
k}g^{kj}_{(1)}
+ \frac{1}{4}\Big( \sigma + \frac{1}{2m^2} \Big)(\Tr\, g_{(1)})g^{ij}_{(1)}  \nn \\
&& + \bigg[ \frac{1}{2}\sigma R_{(0)} -\frac{1}{2m^2}\Tr\, g_{(2)} +
C_1\Tr\, g^2_{(1)} +C_2 (\Tr\, g_{(1)})^2\bigg]g^{ij}_{(0)}\,,\eea
where
\[   C_1 = -\frac{3}{4m^2}B + \frac{3}{m^4}C + \frac{5}{2m^4}D + \frac{1}{2}\Big(\sigma+\frac{5}{8m^2}\Big)  \,, \qquad  C_2 = -\frac{1}{4m^2}B + \frac{2}{m^4}C + \frac{1}{2m^4}D  +\frac{1}{16m^2}   \,. \]
After the conditions for the cancelation of divergences are imposed,
these are given by
\[
 C_1 = \frac{1}{m^4}D + \frac{1}{8}\Big(\sigma + \frac{1}{m^2}\Big)\,, \qquad C_2 = \frac{1}{m^4}C -\frac{1}{8}\sigma\,.
\]
Note that the $\eta$$\eta$-component of EOMs in this case,
$\CE^{\eta}_{~\eta}=0$ gives us the following equations  (See
Appendix B)
\beq \Tr\, g^2_{(1)} = (\Tr\, g_{(1)})^2\,, \qquad R_{(0)} +2\Tr\,
g_{(2)}  -\frac{1}{2}(\Tr\, g_{(1)})^2=0\,.\label{trEOM}\eeq
Using the first equation in the above, one obtains at the special
point of $\sigma=2m^2=1$
\beq \label{rst}
 8\pi G\, T^{ij} =   2 g^{ij}_{(2)} - g^{~~i}_{(1)\, k}g^{kj}_{(1)}  + \frac{1}{2}(\Tr\, g_{(1)})g^{ij}_{(1)}      + \bigg[\frac{1}{2}R_{(0)} -\Tr\, g_{(2)}   +  \Big\{ 4(C+D) + \frac{1}{4} \Big\} (\Tr\, g_{(1)})^2\bigg]g^{ij}_{(0)}\,. \eeq

As can be seen in the above, one  needs another condition to
determine renormalized stress tensor unambiguously.  In fact, by
supposing the validity of $AdS/CFT$ correspondence in this case one
can see that
 there should be another condition given by
\beq
          C+D=0 \,. \label{CondC}\eeq
This condition is accounted for by supposing the consistency of
holographic renormalization  through  correlator computation among
boundary stress tensors, which will be done  in  section 4.
Supposing this condition to hold,
  {\it on-shell} holographically renormalized boundary stress tensor is independent of counter terms and then  the trace part of the renormalized boundary stress tensor is, through the second equation in  Eq.~(\ref{trEOM}),  given by
\beq 8\pi G T =   R_{(0)}    \,. \eeq
As a result, one can see that the central charge of dual CFT in this
case is given by
\beq c = \frac{3}{G}\,, \eeq
which is   twice of the pure Einstein gravity case and  consistent
with Cardy formula~\cite{Oliva:2009ip} or slightly modified central
charge function formalism~\cite{Nam:2010dd}.

One may wonder whether it is possible to consider the following more
general form of counter terms
\[
  S_{c.t.} =     \frac{1}{8\pi G} \int d^2x \sqrt{-\gamma}~(A +B \hat{f} +C\hat{f}^2 +Df_{ij}f^{ij} + Es +Fs^2)\,.
\]
As one can see from EOMs, $s$ terms lead  just to the constant value
when EOM's are imposed, which plays the same role with the boundary
cosmological constant. Therefore, it cannot be used as new counter
terms at all. One may also wonder the possibility of adding even
higher order terms in $\hat{f}$ or $f_{ij}$. These do not lead to
further possibilities, since  these give us the unphysical
ambiguities up to the relevant order in boundary stress tensor. Some
more comments are in order. Our results imply that the cancelation
of divergences are not sufficient to determine the relevant counter
terms completely, which was the case of the pure Einstein gravity.
Even with some additional information from the compliance with
$AdS/CFT$ correspondence, which will be done in section 4,  there is
still ambiguity in counter terms, though this ambiguity does not
affect physical quantities like central charges.

\section{Linearized Analysis and Stress Tensor Correlators}
The linearized analysis in the Fefferman-Graham coordinates is done
in NMG at the so-called critical point $-\sigma=2m^2=1$,  where the
existence of $\log$ modes leads to non-unitary boundary CFT known as
LCFT~\cite{Grumiller:2009sn, Grumiller:2010tj} . In our main
interest, the values of parameters are different   from the critical
point and the analysis should be done independently. In this section
we present the linearized expressions of some quantities and then
represent stress tensor in terms of boundary values of linearized
metric in generic parameter values as well as in the special point
$\sigma=2m^2=1$. Then, we   obtain correlators of stress tensor,
according to the standard $AdS/CFT$ dictionary.

One may note that Fefferman-Graham expansion  contains sufficient
information for the linearized analysis, in some sense.  Once
Fefferman-Graham expansion is obtained,  the linearized analysis can
be implemented easily. For the linearized analysis the background
$AdS$ space can be taken just by the flat boundary metric,
$\eta_{ij}$ with the radial part. Then, the $n (\ge 1)$ order parts
in Fefferman-Graham expansion of two dimensional metric, $\gamma$,
can be regarded as the small fluctuation ones. In addition, the
boundary metric $g_{(0)}$ is also decomposed as the flat one and the
small fluctuation part $h_{(0)}$.  In summary, one may take
\beq g^{(0)}_{ij} = \eta_{ij} + h^{(0)}_{ij}\,, \qquad g^{(n)}_{ij}
= h^{(n)}_{ij}\,, \eeq
and ignore higher order terms for $h$ in the Fefferman-Graham
expanded expressions.

 As a preliminary step, let us consider the pure Einstein gravity case with negative cosmological constant.   Renormalized boundary stress tensor  in terms of linearized metric can be obtained from   the Fefferman-Graham expanded form  as
 \[ 8\pi G\, T^{ij} =  g^{ij}_{(2)} - (\Tr\, g_{(2)})g^{ij}_{(0)}   = g^{ij}_{(2)} + \frac{1}{2}R_{(0)} g^{ij}_{(0)} =h^{ij}_{(2)} + \frac{1}{2}(\p^k\p^lh^{(0)}_{kl} -\p^k\p_k h^{~l}_{(0)\,l})\eta^{ij} +\CO(h^2)\,. \]
By the Wick-rotation of the boundary metric by $\tau=-it$ and
introducing the complex coordinates as   $z=x+ i\tau$, one obtains
the holomorphic  linearized expression of stress tensor as
\[8\pi G\, T_{zz} = h^{(2)}_{zz}\,, \qquad 8\pi G\, T_{\bar{z}\bar{z}} =h^{(2)}_{\bar{z}\bar{z}}\,, \qquad    8\pi G\, T_{z\bar{z}} =  h^{(2)}_{z\bar{z}} + \frac{1}{2}\Big(\p^2h^{(0)}_{\bz\bz} + \bp^2 h^{(0)}_{zz} -2\p\bp h^{(0)}_{z\bz}\Big)\,. \]
To represent (on-shell) stress tensor in terms of boundary values,
one needs to solve EOMs  in terms of boundary values. More
explicitly, one needs to solve $h_{(2)}$ in terms of $h_{(0)}$,
which can be accomplished by linearizing EOMs. Note that $\eta
i$-components of  linearized EOMs  (from the 2nd order part of
Fefferman-Graham expansion) is given by
\beq \bp  h^{(2)}_{zz} = \p h^{(2)}_{z\bz}\,, \qquad   \p
h^{(2)}_{\bz\bz} =  \bp h^{(2)}_{z\bz}\,, \label{LEOMa}\eeq
and $\eta\eta$-component of linearized EOMs is  by
\beq h^{(2)}_{z\bz} = -\frac{1}{2}\Big(\p^2h^{(0)}_{\bz\bz} + \bp^2
h^{(0)}_{zz} -2\p\bp h^{(0)}_{z\bz}\Big)\,, \label{LEOMb}  \eeq
which can be recognized as the linearization of the equation
 $R_{(0)} + 2\Tr\, g_{(2)}=0$.

Through the above linearized EOMs, one obtains linearized on-shell
stress tensor in terms of boundary values as
\beq  T_{zz} = \frac{1}{8\pi G}\frac{\p}{\bp} h^{(2)}_{z\bz} =
-\frac{1}{16\pi G} \frac{\p}{\bp} \Big(\p^2h^{(0)}_{\bz\bz} + \bp^2
h^{(0)}_{zz} -2\p\bp h^{(0)}_{z\bz}\Big) \,, \qquad T_{z\bar{z}}
=0\,. \eeq
Note that the first term in $T_{zz}$ is non-local expression while
the other two are local ones. The standard $AdS/CFT$ dictionary
dictates us the identification $T = \left<T\right>_{CFT}$ and   the
prescription for correlators containing stress tensor as
\[
  \langle T^{ij} \cdots \rangle = \frac{2}{\sqrt{-\det \eta}} \frac{\delta}{\delta h^{(0)}_{ij}} \langle \cdots \rangle \,,
\]
one obtains
\beq \left< T_{zz}(z)T_{zz}(0) \right> = \frac{3}{2G}\frac{1}{8\pi^2
z^4} \equiv \frac{c/2}{4\pi^2 z^4}\,, \eeq
up to irrelevant local expression. There are similar expressions for
anti-holomorphic quantities and correlators between holomorphic and
anti-holomorphic stress tensor vanish. One can  read the central
charge of dual CFT from this formula as $c= 3/2G$.

Now, let us return to our case in NMG. At first sight, one may guess
that one should settle ambiguities in counter terms and should treat
the special  point of $\sigma=2m^2=1$ separately from a generic
parameter point.  On the contrary,  at the linearized level, the
condition of divergence cancelation of Brown-York stress tensor
leads to the following universal form  of renormalized stress tensor
(up to overall scale $\sigma + 1/2m^2$)
\beq  8\pi G\, T^{ij} =2 g^{ij}_{(2)}+ R_{(0)} g^{ij}_{(0)}  = 2
h^{ij}_{(2)}+  (\p^k\p^lh^{(0)}_{kl} -\p^k\p_k
h^{~l}_{(0)\,l})\eta^{ij} +\CO(h^2)\,, \eeq
which also holds at the special point.  
The simple reason of this form is  that the $g_{(1)}=h_{(1)}$ term
appears quadratically in finite part of Brown-York stress tensor. In
some sense  linearization eliminates ambiguities or the details of
counter terms are irrelevant as long as they cancel the divergence
of Brown-York tensor.

It is interesting to  recognize that the linearized expression of
renormalized boundary stress tensor in this case is just twice of
the pure Einstein gravity case up to the given order.   Since
linearized EOMs in NMG are given by the same form in
Eqs.~(\ref{LEOMa}) and~(\ref{LEOMb})  as can be shown  by the
linearization of Fefferman-Graham expansion of EOMs given in the
Appendix B,  one obtains
\beq \left< T_{zz}(z)T_{zz}(0) \right> = \Big(\sigma+
\frac{1}{2m^2}\Big)\frac{3}{2G}\frac{1}{8 \pi^2 z^4}\,, \qquad
\left< T_{zz}(z)T_{\bz\bz}(0) \right> =0\,, \eeq
with similar expression for anti-holomorphic stress tensor. This
gives us the central charge of dual CFT as
\beq c = \Big(\sigma+ \frac{1}{2m^2}\Big)\frac{3}{2G}\,,
\label{Lcentral}\eeq
which holds even at the special point of $\sigma=2m^2=1$. It is also
interesting to observe that $g_{(1)}$ term does not play any role in
the linearized analysis because it appears in  higher order terms
than linear one.  In previous sections, the $g_{(1)}$ term has some
effects in various expressions since we have considered non-linear
Fefferman-Graham expansions. A quadratic $g_{(1)}$ term is also
important requisite for obtaining the correct mass and angular
momentum of new type black holes  through  holographically
renormalized boundary stress tensor as will be shown in the next
section.

One may wonder how NMG, which is a higher derivative theory, can
give us the same linearized EOMs with Einstein gravity.  The answer
is that the magic of Fefferman-Graham expansion is played and
reduces   the radial derivatives effectively in some sense. The
similar phenomenon occurs in the TMG case~\cite{Skenderis:2009nt},
where it was shown that linearized EOMs in TMG can be solved with
some integration constants. It turns out that  the final relevant
equations for correlators are identical with the linearization of
Fefferman-Graham expanded EOMs. That is to say, Fefferman-Graham
expansion  commutes with the linearization.

Now, let us explain why the condition Eq.~(\ref{CondC}) should be
imposed at the special point of $\sigma=2m^2=1$.  In this linearized
approach  the central charge is determined uniquely by the value in
Eq.~(\ref{Lcentral}) just with the condition of divergence
cancelation.     To be consistent with this result, we should impose
the condition~(\ref{CondC}). Without this another condition for
counter terms, the central charges obtained by the trace anomaly of
renormalized stress tensor would be different from those  by stress
tensor correlators. That is to say, the self-consistency in
holographic renormalization governs possible counter terms. As will
be shown in the next section, this another condition is sufficient
to obtain correctly mass and angular momentum of relevant black
holes, {\it i.e.} new type black holes.

\section{Conserved Charges  and Renormalized Stress Tensors}
It is  a  subtle problem to define conserved charges in gravity
theory especially with higher derivative terms. However, in NMG
there are some progress to define mass and angular momentum of $AdS$
black hole solutions. Mass and angular momentum of BTZ black holes
can be easily understood in many ways. Those of new type black holes
are now understood through several studies. Holographically
renormalized stress tensor can also be used to define conserved
charges for black holes. As an application of renormalized stress
tensor, we will derive mass and angular momentum of  BTZ and new
type black holes, which are consistent with the previous results. As
one can see by this computation, it turns out that any counter term
consistent with  holographic renormalization  leads to the correct
mass and angular momentum of black holes, which means that the
remaining ambiguities in counter terms are unphysical. This is also
understood from the fact that the expression (\ref{rbst}) of the
renormalized stress tensor is independent of unphysical ambiguities.
Here, the consistency with holographic renormalization means that
only one condition of Eq.~(\ref{cds}) holds for the  $g_{(1)}=0$
case and three conditions of  Eqs.~(\ref{cds}),~(\ref{cds1})
and~(\ref{CondC}) do for the  $g_{(1)} \neq 0$ case.

It has been known that the mass and the angular momentum of black
holes may be defined in terms of    renormalized stress tensor as
\[ M =  \int^{2\pi}_{0} d\phi~ T^{tt}\,, \qquad J =  \int^{2\pi}_{0} d\phi~ T^{t\phi} \,. \]
As the case of $AdS$ black holes corresponding to the case of
$g_{(1)} =0$,  let us consider the following Schwarzschild form of
BTZ black holes
\[
 ds^2 = \frac{r^2}{(r^2-r^2_+)(r^2-r^2_-)}dr^2  - \frac{(r^2-r^2_+)(r^2-r^2_-)}{r^2}dt^2 + r^2\Big(d\phi  - \frac{r_+r_-}{r^2}dt\Big)^2\,,
\]
which exist at any point of the parameter space. By the coordinate
transformation
\[
 r^2 = e^{2\eta} + \frac{1}{2}(r^2_+ + r^2_-) + \frac{1}{16}e^{-2\eta}(r^2_+ - r^2_-)^2\,,
\]
one obtains  the Fefferman-Graham expanded form for the boundary
metric $\gamma = e^{2\eta}g$ as \beq
 g_{ij} dx^idx^j =  -dt^2 + d\phi^2 + e^{-2\eta}\Big\{ \frac{1}{2}(r^2_+ + r^2_-)(dt^2 + d\phi^2) - 2 r_+r_-dtd\phi\Big\} + \CO(e^{-4\eta})\,,  \nn
\eeq
which shows us explicitly that $g_{(1)}$ vanishes  in this case.
Using renormalized stress tensor given in~(\ref{ST1}) for the
$g_{(1)}=0$ case, one obtains
\beq   M= \frac{r^2_+ +r^2_-}{8G}\Big(\sigma+\frac{1}{2m^2}\Big)\,,
\qquad J = \frac{r_+r_-}{4G}\Big(\sigma+\frac{1}{2m^2}\Big) \,. \eeq
These are consistent with  previous results by various other
methods~\cite{Clement:2009gq, Nam:2010ma, Nam:2010dd}.

Now, let us consider new type black holes which exist at the special
point $\sigma=2m^2=1$. Since Fefferman-Graham expansion can be  done
completely for static new type black holes, rotating case will be
dealt with separately. The metric of static new type black holes is
given by
\beq\label{nmg-metric}
  ds^2 =   -(r^2 + b r + c)dt^2 + \frac{dr^2}{(r^2 + b r + c)} + r^2 d\phi^2
   \,,
\eeq
with  outer and inner horizons  at $r_{\pm} = \half (-b \pm
\sqrt{b^2-4c})$. The mass of static new type black holes is
identified as a conserved charge by~\cite{Nam:2010ma}
\beq M = \frac{b^2-4c}{16G} \,, \eeq
which is also justified by the dynamical approach~\cite{Maeda:2010}
or $AdS/CFT$ correspondence~\cite{Nam:2010dd}.
%
%
By the following coordinate transformation
\beq  r \equiv e^{\eta} - \frac{b}{2} +
\frac{1}{4}e^{-\eta}\Big(\frac{b^2}{4} - c \Big)\,, \eeq
one can see that the relevant part of the  metric can be put in the
Fefferman-Graham expanded form as
\bea  g_{ij}dx^idx^j &=& - \bigg[ 1 - \frac{1}{4}e^{-2\eta}\Big(\frac{b^2}{4}-c\Big)\bigg]^2\, dt^2 + \bigg[ 1 - \frac{b}{2}e^{-\eta} +\frac{1}{4}e^{-2\eta}\Big(\frac{b^2}{4}-c\Big)\bigg]^2\, d\phi^2 \nn \\
&=& -dt^2 + d\phi^2 - be^{-\eta} d\phi^2 +
\frac{1}{2}e^{-2\eta}\bigg[ \Big(\frac{1}{4}b^2-c\Big)dt^2 +
\Big(\frac{3}{4}b^2-c\Big) d\phi^2\bigg] +  \CO(e^{-3\eta})\,.  \nn
\eea
Using renormalized stress tensor relevant for the special point of
$\sigma=2m^2=1$ given in Eq.~(\ref{rst}) equipped with
Eq.~(\ref{CondC}), one obtains
%
%
%
%
%
\beq 8\pi G\, T^{tt} = 2g^{tt}_{(2)}  +\Tr\, g_{(2)}
-\frac{1}{4}(\Tr\, g_{(1)})^2 = \frac{1}{4}(b^2-4c)\,.\eeq
One can see that this gives   the correct mass of static new type
black holes~\cite{Giribet:2010ed, Nam:2010ma, Maeda:2010} .

Now, let us consider the rotating new type black holes
\cite{Oliva:2009ip, Giribet:2009qz}.
\beq ds^2 = -N^2(r)F^2(r)dt^2 + \frac{dr^2}{F^2(r)} +r^2\Big(d\phi
+N^{\phi}(r)dt \Big)^2\,, \eeq
where
\bea N(r) &\equiv& 1 + \frac{b\alpha}{2H(r)}\,, \qquad N^{\phi}(r) \equiv  \frac{\sqrt{\alpha(1-\alpha)}}{r^2}\Big(c + bH(r)\Big)\,, \nn \\
       F(r) &\equiv & \frac{H(r)}{r}\bigg[H^2(r)  + b(1-\alpha) H(r) + \frac{b^2}{4}\alpha^2 +c (1-2\alpha)\bigg]^{1/2}\,,  \nn \\
       H(r) &\equiv & \bigg[ r^2 -\frac{b^2}{4}\alpha^2 + c\, \alpha\bigg]^{1/2}\,. \nn \eea
Note that $\alpha$ is the parameter for rotation $(0\le \alpha \le
1/2)$, which vanishes in the static case. In the following, one can
see the explicit formula for the mass and the angular momentum of
rotating new type black holes and so the relation of the parameter
$\alpha$ with the mass and the angular momentum. The above metric
can be set in the appropriate form for the Fefferman-Graham
expansion at least up to relevant order by the following ansatz
\[  r  = e^{\eta}  + \beta_1+ \beta_2 e^{-\eta} +\CO(e^{-2\eta} )\,. \]
Using the coordinate transformation dictated by $dr/F(r) = d\eta$,
one obtains
\[  \beta_1 \equiv  -\frac{b}{2}(1-\alpha) \,, \qquad \beta_2 \equiv  \frac{1}{4}\Big[  \frac{b^2}{4}(1-2\alpha +2\alpha^2) -  c   \Big] \,.\]
Then, one can see that
\bea
 g_{tt} &=&  -1 - b\alpha\, e^{-\eta} - \Big[b\beta_1+\beta^2_1 + 2\beta_2  +c\Big]e^{-2\eta} + \CO(e^{-3\eta}) \,, \nn \\
g_{\phi \phi} &=& 1  - b(1-\alpha)\, e^{-\eta} + (\beta^2_1 + 2\beta_2)  e^{-2\eta}+ \CO(e^{-3\eta})\,, \nn \\
g_{t\phi} & =& b\sqrt{\alpha(1-\alpha)}\, e^{-\eta}+
\sqrt{\alpha(1-\alpha)} (b\beta_1+c)  e^{-2\eta}+ \CO(e^{-3\eta})\,,
\nn \eea
%
%
%
which gives  the correct mass and angular momentum of rotating new
type black holes irrespective of the choice of counter terms with
which the three conditions  should be satisfied. Explicitly, one
obtains
\bea 8\pi G\, T^{tt} &=& 2g^{~tt}_{(2)} - g^{2~ tt}_{(1)} +\frac{1}{2}(\Tr\, g_{(1)}) g^{~ tt}_{(1)} + \Tr\, g_{(2)} - \frac{1}{4}(\Tr\, g_{(1)})^2 = \frac{b^2}{4}-c\,,  \\
8\pi G\, T^{t\phi} &=& 2g^{~t\phi}_{(2)} - g^{2~ t\phi}_{(1)}
+\frac{1}{2}(\Tr\, g_{(1)}) g^{~ t\phi}_{(1)}  =
\Big(\frac{b^2}{4}-c\Big)2\sqrt{\alpha(1-\alpha)}\,. \eea
Therefore, one can see that the mass and the ratio of the mass and
the angular momentum is given respectively by
\beq M = \frac{1}{16G}(b^2-4c)\,, \qquad \frac{J}{M} =
2\sqrt{\alpha(1-\alpha)}\,. \eeq
One may note that new type black holes become BTZ ones when the
black hole parameter $b$ vanishes and in this case the mass and the
angular momentum of new type black holes also become those of BTZ
black holes. The lesson of the above computation is that there is
one parameter family of consistent counter terms and it leads to the
correct mass and angular momentum of rotating new type black holes
as well as central charges.

\section{Conclusion}
In this paper we have adopted Fefferman-Graham expansion in NMG and
studied holographic renormalization. We have particularly focused on
the special parameter point given by $\sigma =2m^2 =1$. At this
special point the fall-off boundary condition is weaker than
Brown-Henneaux one and so it needs to be analyzed separately. After
obtaining renormalized stress tensor in generic Fefferman-Graham
expansion, we obtain the central charge of dual CFT and then confirm
mass and angular momentum of  (rotating) new type black holes in
this formalism. We have also solved the linearized EOMs in
Fefferman-Graham expansion and then obtained correlators of
renormalized boundary stress tensor. The central charge of dual CFT
can also be read from these correlators.

To obtain renormalized stress tensor, we have considered   generic
counter terms and showed that there are some ambiguities in their
construction, while physical quantities are innocent of these
ambiguities. This is a reminiscent of renormalization scheme
independence in field theories. However, it is not clear at this
stage that this should really be understood as corresponding to
scheme independence. The previous prescription for counter terms can
be understood as  minimal choices of those in our perspective.

Some comments are in order. At a generic point in the parameter
space of the NMG Lagrangian, our expansion with $g_{(1)}=0$ is a
rather generic one and consistent with the fall-off conditions of
the pure Einstein case (See Appendices B and C). In this case
ambiguities in counter terms are rather large, while these
ambiguities are unphysical. At the special point in our interest,
our expansion is also consistent as shown
in~\cite{Oliva:2009ip}\cite{Giribet:2010ed} and  it satisfies  a
weaker fall-off condition than the Brown-Henneaux one. Our results
show that this fall-off condition is consistent  with holographic
renormalization with ambiguities in counter terms. In fact, those
ambiguities are shown to be physically irrelevant.

Now, it is useful to comment on our results  in minds for the
generality of our Fefferman-Graham expansion. Since we are dealing
with a higher curvature or derivative theory, there is no guarantee
that our expansion form is the most generic one, while  the generic
expansion in the pure Einstein case was derived rigorously
in~\cite{FG}.  Our observation of the  ambiguities in counter terms
may be originated from the lack of generality of the
Fefferman-Graham expansion.  With the most generic expansion, it
might be possible to determine the counter terms completely.
Explicitly, when a  more generic expansion is considered  there may
be other power law or `$\log$' terms leading to new divergence
structure. To render finite new divergence structure, additional
information about counter terms may be obtained and then the counter
terms may be determined completely.   However, even with the lack of
the generality proof for the expansion  our results are consistent
and can be interpreted as the consistency of holographic
renormalization  with the restricted form of Fefferman-Graham
expansion, which is reminiscent of the consistency  claim of  chiral
gravity without `$\log$' modes. In other words,  holographic
renormalization for higher derivative theories has no inconsistency
with a certain truncation.

Notably, it has been known that at the special point in our interest
a more generic expansion is possible. That is to say, there may be
the so-called `$\log$' terms, which is forbidden in a generic point
by equations of motions, in the form of
\[  \gamma_{ij}  =  e^{2\eta} g^{(0)}_{ij} + e^{\eta}  g^{(1)}_{ij}+ \eta\, e^{\eta}  b^{(1)}_{ij}  +  g^{(2)}_{ij} +  e^{-\eta} g^{(3)}_{ij} +   e^{-2\eta}g^{(4)}_{ij} + \CO(e^{-3\eta})\,.  \]
This `$\log$' fall-off condition is also shown to be consistent
one~\cite{Oliva:2009ip}\cite{Grumiller:2010tj}. With this more
generic expansion, one may also study holographic renormalization
and determine counter terms completely. However, it should be noted
that this expansion satisfies different fall-off conditions and
needs to be treated separately. One may also note that another
$`\log$' term, $\eta  b_{(2)}$, exists at the critical point which
is different from the above $`\log$' term $\eta e^{\eta} b_{(1)}$,
which needs to be investigated independently, as well.

Though there are some ambiguities in the choice of counter terms in
our approach,  it will be very interesting to investigate the
possibility of  determining counter terms completely at any point of
parameter space through  more generic, yet unknown, Fefferman-Graham
expansion   and/or  other formalism. There have been some approaches
to determine counter terms uniquely in the pure Einstein or
Gauss-Bonnet gravity~\cite{de
Boer:1999xf}\cite{Papadimitriou:2004ap}\cite{Liu:2008zf}\cite{Grumiller:2009dx}.
It will be very interesting to  elaborate more on these methods for
higher derivative gravity theories like NMG.

At the linearized analysis  the existence of $g_{(1)}$ at the
special point of $\sigma=2m^2=1$ does not have a crucial role, since
it does not lead to modification to EOMs of $h_{(2)}$ compared to
the pure Einstein case. This situation is very similar to the case
of $\log$ modes at the critical point. At the critical point, the
existence of $\log$ modes leads to $AdS/LCFT$ correspondence, which
is related to the introduction of new boundary source terms in the
scheme of holographic renormalization,  though the $\log$ fall-off
behavior of bulk modes can be analyzed without their boundary source
terms.  Since these new boundary terms break asymptotic $AdS$
properties, these are set to be zero at the end of computation,
while their existence leads to new anomaly in LCFT, known as $b$
central charge.  It may be straightforward to apply such procedure
to our case. However, it is unclear what is the dual interpretation
of these new source terms in our case, since it is unclear which CFT
corresponds to this case. It would be very interesting to see
whether there is dual CFT with the inclusion of source terms for
$g_{(k)}$ part of odd $k$  in the bulk metric modes.

Though we have shown that Fefferman-Graham expansion reduces the
linearized EOMs in NMG to those in the Einstein gravity case, it
will be interesting to verify this statement by direct computation
from linearized EOMs. By solving linearized EOMs with higher
derivatives, one encounters several integration constants. After
all, these constants should not play significant roles to obtain
correlators for asymptotic $AdS$ space as in
TMG~\cite{Skenderis:2009nt}. Another interesting problem is to apply
holographic renormalization to extended NMGs. Since it was shown
that new type black holes exist in $R^3$-NMG~\cite{Nam:2010dd}, it
is clear that special points also exist in $R^3$-NMG.  Since the
parameter space is also extended  in extended NMG, it is expected
that there are several special subspaces. Therefore, it will be
interesting to study more extensively the parameter space and see
the role of holographic renormalization in these theories.

\section*{Acknowledgements}


S.N and S.H.Y were supported by the National Research Foundation of
Korea(NRF) grant funded by the Korea government(MEST) through the
Center for Quantum Spacetime(CQUeST) of Sogang University with grant
number 2005-0049409. S.H.Y would like to thank Prof. Seungjoon Hyun
at Yonsei University for discussion some time ago. S.N and J.D.P
were supported by a grant from the Kyung Hee University in
2009(KHU-20110060). S.N and Y.K were supported by Basic Science
Research Program through the National Research Foundation of
Korea(NRF) funded by the Ministry of Education, Science and
Technology(No.2011-0004328). The authors would like to thank referee
for valuable comments and also thank S. Pal for some discussion.

\newpage
\appendix
 \renewcommand{\theequation}{A.\arabic{equation}}
  \setcounter{equation}{0}
\section{Useful Formulae}

In this appendix, we present  various useful formulae in
Fefferman-Graham coordinates for various geometrical quantities such
as metric, extrinsic curvature and Ricci tensor.  These formulae are
used in the main text in order to compute Brown-York stress tensor
and renormalized one.

The Fefferman-Graham expansion of the metric given in~(\ref{FGexp})
leads to
\bea \gamma^{i}_{\,j} &=& g^{~i}_{(0)j} = \eta^{i}_{\,j}\,, \nn \\
\gamma^{ij} &=& e^{-2\eta} g_{(0)}^{ij} -  e^{-3\eta}  g_{(1)}^{ij}
- e^{-4\eta} \left( g_{(2)}^{ij} - (g^2_{(1)})^{ij}\right)
\nn \\
&& - e^{-5\eta} \left( g^{ij}_{(3)} - (g_{(1)}g_{(2)})^{ij} -
(g_{(2)}g_{(1)})^{ij} + (g_{(1)}^3)^{ij}\right)   + \CO(e^{-6\eta})  \nn \\
\sqrt{-\gamma} &=& e^{2\eta}\sqrt{-g_{(0)}}\bigg\{ 1 + \half
e^{-\eta}\Tr\, g_{(1)} + \half e^{-2\eta}\Big[\Tr\, g_{(2)} -\half
\Tr\, g^{2}_{(1)}  + \frac{1}{4}(\Tr\, g_{(1)})^2 \Big]  \nn \\
&& \!\!\!\!\!  + \half e^{-3\eta}\Big[\Tr\, g_{(3)}
-\Tr(g_{(1)}g_{(2)}) + \frac{1}{3}\Tr\,g^3_{(1)} + \half \Big(\Tr\,
g_{(2)} -\frac{1}{2}\Tr\, g^2_{(1)} + \frac{1}{12}(\Tr\, g_{(1)})^2
\Big)\Tr\, g_{(1)}\Big]\nn \\
&& \!\!\!\!\!  + \CO(e^{-4\eta})  \bigg\}\,, \nn \eea
where in the right hand side of equalities all indices are raised or
lowered by $g^{(0)}$ not $\gamma$ and $\Tr$ denotes the contraction
with $g_{(0)}$. Christoffel symbols for the boundary metric are
given  by
\bea \Gamma^{k}_{ij} &=& \Gamma^{~~k}_{(0) \, ij} + \half
e^{-\eta}\,g^{kl}_{(0)}\Big(\bar{\nabla}_{i}g^{(1)}_{lj} +
\bar{\nabla}_{j}g^{(1)}_{il} -\bar{\nabla}_{l}g^{(1)}_{ij}\Big) \nn
\\
&&    + \half  e^{-2\eta}\bigg[  g^{kl}_{(0)}
\Big(\bar{\nabla}_{i}g^{(2)}_{lj} + \bar{\nabla}_{j}g^{(2)}_{il}
-\bar{\nabla}_{l}g^{(2)}_{ij}\Big)   -g^{kl}_{(1)}
\Big(\bar{\nabla}_{i}g^{(1)}_{lj} + \bar{\nabla}_{j}g^{(1)}_{il}
-\bar{\nabla}_{l}g^{(1)}_{ij}\Big) \bigg]   + \CO(e^{-3\eta})
\nn \\
&\equiv &  \Gamma^{~~k}_{(0) \, ij} +  e^{-\eta}\, \Gamma^{~~k}_{(1)
\, ij} +  e^{-2\eta}\,\Gamma^{~~k}_{(2) \,ij}+
 \CO(e^{-3\eta})\,.
 \label{CCon}
  \eea
where $\bar{\nabla}$ denotes the covariant derivative with respect
to $g_{(0)}$.

The Fefferman-Graham expansion of extrinsic curvature tensor is
given by
\bea K_{ij} &=& e^{2\eta} g^{(0)}_{ij} + \half e^{\eta} g^{(1)}_{ij}
-\half e^{-\eta} g^{(3)}_{ij}  -e^{-2\eta}g^{(4)}_{ij} +
\CO(e^{-3\eta})\,,  \nn \\
K^{i}_{\, j} &=& \eta^{i}_{\, j} - \half e^{-\eta}  g^{~ i}_{(1)j} -
e^{-2\eta}\Big[g^{~i}_{(2)j} -\half (g^2_{(1)})^{i}_{\, j} \Big] \nn \\
&& - e^{-3\eta}\Big[ \frac{3}{2}g^{~~ i}_{(3)\, j} -
(g_{(1)}g_{(2)})^{i}_{\, j} - \half (g_{(2)}g_{(1)})^{i}_{\, j} +
\half (g^3_{(1)})^{i}_{\, j}
\Big] \nn \\
&& - e^{-4\eta}\bigg[2 g^{~~ i}_{(4)\, j}
-\frac{3}{2}(g_{(1)}g_{(3)})^{i}_{\, j}
-\frac{1}{2}(g_{(3)}g_{(1)})^{i}_{\, j} - (g^2_{(2)})^{i}_{\, j}+
(g^2_{(1)} g_{(2)})^{i}_{\, j}  \nn \\
&&~~~~~ + \frac{1}{2}(g_{(1)}g_{(2)}g_{(1)})^{i}_{\, j} +
\frac{1}{2}(g_{(2)}g_{(1)})^{i}_{\, j} -
\frac{1}{2}(g^4_{(1)})^{i}_{\, j}     \bigg] + \CO(e^{-4\eta})\,,
\nn \eea
%
%
which leads to
\bea &&\sqrt{-\gamma}\Big[(K-1)\gamma^{ij} -K^{ij}\Big] =
\sqrt{-g_{(0)}}\bigg[\half\, e^{-\eta}\Big\{g^{ij}_{(1)} - (\Tr\,
g_{(1)} )g^{ij}_{(0)} \Big\}   + e^{-2\eta}\Big\{ g^{ij}_{(2)}
-(\Tr\, g_{(2)})g^{ij}_{(0)}  \nn   \\
&& ~~~~~~~~~~~ \qquad \qquad
-g^{ik}_{(1)}g^{~~~j}_{(1)k}+\frac{3}{4}(\Tr\, g_{(1)} )
g^{ij}_{(1)} + \half \Big(\Tr\, (g_{(1)})^2 - \frac{1}{2} (\Tr\,
g_{(1)})^2 \Big) g^{ij}_{(0)}\Big\} + \cdots \bigg]\,. \nn \eea

Fefferman-Graham expansion for auxiliary fields $f^{\mu\nu}$'s from
(\ref{auxliary}) are given by
\bea   m^2 f^{ij}  &=&  - e^{-2\eta} g^{ij}_{(0)}  +  2e^{-3\eta}\,   g^{ij}_{(1)}   + e^{-4\eta} \bigg[  g^{ij}_{(2)} - 2g^{i}_{(1)\,k}g^{kj}_{(1)} - \half (\Tr\, g_{(1)})\, g^{ij}_{(1)}  \nn \\
&&~~~~~~~~ +2\Big(R^{ij}_{(0)} -\frac{1}{4}R_{(0)}g^{ij}_{(0)}\Big)  + \Big( \Tr\, g_{(2)}  -\frac{3}{8}\Tr\, g^2_{(1)} + \frac{1}{8}(\Tr\, g_{(1)})^2\Big)\, g^{ij}_{(0)}\bigg] + \CO(e^{-5\eta})\,, \nn \\
m^2f^{i}_{\, j} &=& -\eta^{i}_{\, j} +e^{-\eta}\, g^{~i}_{(1)\, j} + \half\, e^{-2\eta}\bigg[ \Big( R_{(0)} + 2\Tr\, g_{(2)} -\frac{3}{4}\, \Tr\, g^2_{(1)} + \frac{1}{4}(\Tr\, g_{(1)})^2\Big)\eta^{i}_{\, j}  \nn \\
&&~~~~~~~~~~~~~~~~~~~~ -  (\Tr\, g_{(1)})\, g^{~ i}_{(1)\, j} \bigg] + \CO(e^{-3\eta})\,, \nn \\
 m^2 h^{i} &=& e^{-3\eta} \Big[\bar{\nabla}^{i} \Tr\, g_{(1)} -\bar{\nabla}_j g^{ji}_{(1)} \Big] +   2e^{-4\eta} \bigg[\bar{\nabla}^i\Tr\, g_{(2)} -\bar{\nabla}_{j}g^{ji}_{(2)}  \nn \\
 && \qquad \qquad + g^{i}_{(1)j}\bar{\nabla}_kg^{kj}_{(1)}  +\half g^{j}_{(1)k}\bar{\nabla}_{j}g^{ik}_{(1)} -\frac{3}{4}g^{kl}_{(1)}\bar{\nabla}^ig^{(1)}_{kl} -\frac{3}{4}g^{ij}_{(1)}\bar{\nabla}_{j}\Tr\, g_{(1)}\bigg]  + \CO(e^{-5\eta}) \,, \nn \\
 m^2s&=& -1 - \half e^{-2\eta}\Big[R_{(0)} + 2\Tr\, g_{(2)}  - \frac{1}{4}\Tr\, g_{(1)}^2 - \frac{1}{4} (\Tr\, g_{(1)} )^2 \Big] +\cdots  \,.\eea
Recall that $\hat{f} \equiv \gamma^{ij}f_{ij}$, then
\[ m^2\hat{f} = -2  +   e^{-\eta}\,  \Tr\, g_{(1)} + e^{-2\eta}\bigg[R_{(0)} + 2\Tr\, g_{(2)} -\frac{3}{4}\Tr\, g^2_{(1)} -\frac{1}{4}(\Tr\, g_{(1)})^2\bigg] + \CO(e^{-3\eta}) \,. \]
Note that the original $f$ is given by $f=\hat{f} + s$ as
\[ m^2 f = \frac{1}{2}R = -3 +    e^{-\eta}\,  \Tr\, g_{(1)}  + \half  e^{-2\eta}\bigg[R_{(0)} +  2\Tr\, g_{(2)}   -\frac{5}{4}\Tr\, g^{2}_{(1)} -   \frac{1}{4}  (\Tr\, g_{(1)})^2\bigg]  + \CO(e^{-3\eta}) \,. \]
Some useful formulae for the computation of Brown-York tensor are
\begin{eqnarray*}
 - \nabla^{\left(i\right.} h^{\left.j\right)}
       + \gamma^{ij} \nabla_k h^k
   &=& \CO(e^{-5\eta})  \\
\!\! \frac{1}{2} D_{\eta} f^{ij} + K_k^{\left(i\right.}
       f^{\left.j\right)k} - \frac{1}{2}\gamma^{ij} D_{\eta} f
   &=& - \frac{1}{2m^2} e^{-3\eta} \Big( g_{(1)}^{ij}
       - \Tr\,g_{(1)} g_{(0)}^{ij} \Big)
       + \frac{1}{2m^2}e^{-4\eta} \bigg[ (g_{(1)}^2)^{ij}
       + R_{(0)} g_{(0)}^{ij}  \\
   &&  + \Big( 2 \Tr\,g_{(2)}
       - \frac{3}{4}(\Tr\, g_{(1)})^2 - \frac{3}{4} \Tr\, g_{(1)}^2
       \Big) g_{(0)}^{ij} \bigg] + \CO(e^{-5\eta})
\end{eqnarray*}
Useful formulae for counter terms computation are
\begin{eqnarray*}
 m^4 \hat{f}^2
    &=& 4 - 4 e^{-\eta} \Tr\, g_{(1)}
        - 4 e^{-2\eta} \Big[ R_{(0)} + 2 \Tr\, g_{(2)}
        - \frac{3}{4} \Tr\, g_{(1)}^2
        - \frac{1}{2} (\Tr\, g_{(1)})^2 \Big]
        + \CO(e^{-3\eta})  \\
 m^4 f_{kl} f^{kl}
    &=& 2 - 2 e^{-\eta} \Tr\, g_{(1)}
        - 2 e^{-2\eta} \Big[ R_{(0)} + 2 \Tr\, g_{(2)}
        - \frac{5}{4} \Tr\, g_{(1)}^2
        - \frac{1}{4}(\Tr\, g_{(1)})^2 \Big]
        + \CO(e^{-3\eta})
\end{eqnarray*}

Fefferman-Graham expansion for Ricci tensor and Ricci scalar is
given by
\bea
 R^{i}_{\, j} &=& -2\eta^{i}_{j} +\half
e^{-\eta}\Big[g^{i}_{(1)j}
+ (\Tr\, g_{(1)}) \eta^{i}_{j} \Big] \nn \\
&& + e^{-2\eta}\Big[ \half\Big( R_{(0)}+  2\Tr\, g_{(2)} - \Tr\,
g^{2}_{(1)} \Big) \eta^{i}_{j} -\frac{1}{4}(\Tr\, g_{(1)})
g^{i}_{(1)\, j}  \Big]   \nn \\
&& + \half e^{-3\eta}\bigg[  - R_{(0)} g^{~~ i}_{(1)\, j} + 2
X^{i}_{j}  +    \Big( 3\Tr\, g_{(3)} -3 \Tr( g_{(1)}g_{(2)})+ \Tr\,
g^3_{(1)}\Big)\eta^{i}_{\, j}  \nn \\
&&~~~\quad \qquad    -  \Big(3g^{~~ i}_{(3)\, j} -
2(g_{(1)}g_{(2)})^{i}_{\, j}
- (g_{(2)}g_{(1)})^{i}_{\, j}   +  (g_{(1)}^3)^{i}_{\, j} \Big) \nn \\
&& ~~~\quad \qquad    -  \Big( g^{~~ i}_{(2)\, j} - \half
(g_{(1)}^2)^{i}_{\, j}\Big) \Tr\, g_{(1)} - \Big(\Tr\, g_{(2)}
-\half \Tr\, g^2_{(1)}\Big)g^{~~ i}_{(1)\, j}
\bigg] \nn \\
&& + e^{-4\eta}\bigg[ Y^{i}_{\, j} -\left(g_{(1)} X \right)^{i}_{j}
-\frac{1}{2}R_{(0)}\Big(g^{~~
i}_{(2)\, j} -(g_{(1)}^2)^{i}_{\, j}\Big) \nn \\
&& ~~~ \qquad  -  4g^{~~ i}_{(4)\, j} + 3(g_{(1)}g_{(3)})^{i}_{\, j}
+ (g_{(3)}g_{(1)})^{i}_{\, j} +2(g^2_{(2)})^{i}_{\, j}
-2(g^2_{(1)}g_{(2)})^{i}_{\, J} - (g_{(1)}g_{(2)}g_{(1)})^{i}_{\, j}
\nn \\
&& ~~~ \qquad - (g_{(2)}g^2_{(1)})^{i}_{\, j} + (g^4_{(1)})^{i}_{j}
- \frac{1}{4}\Big(3g_{(3)} -2g_{(1)}g_{(2)} -  g_{(2)}g_{(1)} +
g^3_{(1)}\Big)^{i}_{\, j}\Tr\, g_{(1)}  \bigg] \nn \\
&& ~~~ \qquad  -\Big(\Tr\, g_{(2)} -\frac{1}{2}\Tr\,
g^2_{(1)}\Big)\Big(g_{(2)} -\frac{1}{2}g^2_{(1)}\Big)^{i}_{\,j} -
\frac{1}{4}\Big(3\Tr\, g_{(3)} -  3\Tr(g_{(1)}g_{(2)}) + \Tr\,
g^3_{(1)} \Big)g^{~~i}_{(1)\, j} \nn \\
&& ~~~ \qquad +\Big(2\Tr\, g_{(4)} -2\Tr(g_{(1)}g_{(3)}) - \Tr\,
g^2_{(2)} +2\Tr (g^2_{(1)}g_{(2)}) -\frac{1}{2}\Tr\,
g^4_{(1)}\Big)\eta^{i}_{\, j}\bigg] \nn \\
&& + \CO(e^{-5\eta})\,, \nn
          \eea
\bea    R^{\eta}_{~\eta} &=& -2 + \half e^{-\eta}\, \Tr\, g_{(1)}
-\frac{1}{4} e^{-2\eta}\, \Tr\, g^{2}_{(1)} - \half  e^{-3\eta}\Big[
3\Tr\, g_{(3)} -  \Tr( g_{(1)}g_{(2)})\Big]  \nn \\
&& -e^{-4\eta}\bigg[ 4\Tr\, g_{(4)} - \frac{5}{2}\Tr(g_{(1)}g_{(3)})
- \Tr\, g^2_{(2)} + \frac{3}{2} \Tr(g^2_{(1)}g_{(2)})
-\frac{1}{4}\Tr\, g^4_{(1)}\bigg]
+\CO(\eta^{-5\eta})\,, \nn \\
R^{\eta}_{~ i} &=& \frac{e^{-\eta}}{2}  \bigg[\bar{\nabla}_{i}\Tr
g_{(1)} -\bar{\nabla}_{j}g^{j}_{(1)\, i} \bigg]  + e^{-2\eta}\bigg[
\bar{\nabla}_{i}\Tr g_{(2)} -\bar{\nabla}_{j}g^{j}_{(2)\, i}   \nn
\\
&& ~~~~~~~  \qquad + \half g^{(1)}_{ij}\bar{\nabla}_{k}g^{jk}_{(1)}
+ \half g^{jk}_{(1)} \bar{\nabla}_{j} g^{(1)}_{ki}   -\frac{3}{4}
g^{kl}_{(1)} \bar{\nabla}_{i} g^{(1)}_{kl} - \frac{1}{4}g
^{(1)j}_{i}\bar{\nabla}_j\Tr\, g_{(1)} \bigg]+\CO(\eta^{-3\eta})\,,
\nn \\
R &=& -6 +2e^{-\eta}\Tr\, g_{(1)} + e^{-2\eta}\Big[R_{(0)}+2\Tr\, g_{(2)}  -  \frac{5}{4}\, \Tr\, g_{(1)}^2   -\frac{1}{4}(\Tr\, g_{(1)})^2  \Big]  \nn  \\
 && + e^{-3\eta}\bigg[ \Tr\,X -\Tr( g_{(1)}g_{(2)})  + \half \Tr\, g^3_{(1)}   - \half  \Big(R_{(0)} +2   \Tr\, g_{(2)} - \Tr\, g^2_{(1)}  \Big) \Tr\, g_{(1)}
         \bigg]   \nn \\
&&+ e^{-4\eta}\bigg[\Tr\, Y  - \Tr\,(g_{(1)} X)  -4\Tr\, g_{(4)} + \Tr\, g^2_{(2)}    + \frac{5}{2}\Tr(g_{(1)}g_{(3)})   - \frac{3}{2}\Tr(g^2_{(1)}g_{(2)})  +\frac{1}{4}\Tr\, g^4_{(1)}  \nn \\
&&~~~~~    -\frac{1}{2}\Big(R_{(0)} + 2\Tr\, g_{(2)} \Big)\Big(\Tr\, g_{(2)} - \Tr\, g^2_{(1)} \Big)    - \frac{1}{4}(\Tr\, g^2_{(1)} )^2    \nn \\
    &&  ~~~~~     -  \Big( \frac{3}{2}\Tr\, g_{(3)} -\frac{3}{2}\Tr(g_{(1)}g_{(2)}) + \frac{1}{2} \Tr\, g^3_{(1)}\Big)\Tr\, g_{(1)}    \bigg]
     + \CO(e^{-5\eta})\,, \nn \eea
where $X, Y$ are defined in terms of ($\ref{CCon}$) by
\begin{eqnarray}
 X^{i}_{\, j}  &\equiv &  {1 \over 2} \Big(\bar{\nabla}^{k}\bar{\nabla}^{i} g_{(1)\, jk} +
\bar{\nabla}^{k}\bar{\nabla}_{j} g^{~~ i}_{(1)\, k}
-\bar{\nabla}^{i}\bar{\nabla}_{j}\Tr\, g_{(1)}   -\bar{\nabla}^2
g^{~~ i}_{(1)\,j}\Big)  \, , \nn \\
 Y^{i}_{\, j}  &\equiv & g^{ik}_{(0)}\Big(\bar{\nabla}_{l} \Gamma^{l}_{(2)\, jk}-\bar{\nabla}_{j}\Gamma^{~~ l}_{(2)\, lk} + \Gamma^{~~ l}_{(1)\, lm}\Gamma^{~~ m}_{(1)\, jk}- \Gamma^{~~ l}_{(1)\, jm}\Gamma^{~~ m}_{(1)\,
 lk}\Big)\,.
\end{eqnarray}
%
%
%
%

\section{Equations of Motions in Fefferman-Graham Coordinates}

 In this appendix we present EOMs in the Fefferman-Graham expanded form, through which  the relevant  linearized EOMs can be obtained.
Before presenting the expression for $\CE^{\mu}_{~\nu}$,  let us
consider the Fefferman-Graham expansion of the $ \CK^{\mu}_{\,\nu}$
term, which are given by
\begin{eqnarray}
\CK^{\eta}_{\, \eta} &=&  - \frac{1}{2} + {1 \over 2}\,e^{-\eta}\,
\Tr\,g_{(1)} +e^{-2 \eta} \left[ {1 \over 2} R_{(0)} +\Tr\,g_{(2)} -
{1 \over 8} \Tr\,g^2_{(1)} -{3 \over 8} \left( \Tr\,g_{(1)}
\right)^2 \right] \nonumber \\
&& + e^{-3 \eta} \Biggl[ {1 \over 2} \left(
\bar{\nabla}^{i}\bar{\nabla}^{j} g_{(1)\, ij}
-\bar{\nabla}^2\Tr\,g_{(1)} \right)- {1 \over 2} R_{(0)}
\Tr\,g_{(1)} -\Tr\,(g_{(1)}\,g_{(2)}) -{9 \over 16} \Tr\,g_{(1)}
\,\Tr\,g^2_{(1)} \nonumber \\
&& \qquad ~ \quad  - \Tr\,g_{(1)}\, \Tr\,g_{(2)}  + {3 \over 2}
\Tr\,g_{(3)} + {3 \over 4} \Tr\,g^3_{(1)} +{7 \over 16}
(\Tr\,g_{(1)} )^3  \Biggr] +\CO(e^{-4 \eta})\,,
\\
\CK^{\eta}_{\,i} &=& -\frac{e^{-\eta}}{2}  \bigg[\bar{\nabla}_{i}\Tr
g_{(1)} -\bar{\nabla}_{j}g^{j}_{(1)\, i} \bigg]  + e^{-2 \eta}
\bigg[ \bar{\nabla}_{i}\Tr g_{(2)} -\bar{\nabla}_{j}g^{j}_{(2)\, i}
+\CO(g^2_{(1)}) \bigg] +  \CO(e^{-3\eta})\,,
\\
\CK^{i}_{\, j} &=& -{1 \over 2}\,  \eta^{i}_{j} +e^{-\eta} \left[ {1
\over 2} \,(\Tr\,g_{(1)}) \,\eta^{i}_{j}  - {1 \over 2} g^{i}_{(1)
j} \right] \nonumber \\
&& + e^{- 2\eta} \Bigg[ R^{i}_{(0) j} +\bigg\{ -{1 \over 2} R_{(0)}
-{1 \over 8} \left( (\Tr\,g_{(1)} )^2 +\Tr\,g^2_{(1)} \right)
\bigg\}  \eta^{i}_{j}  +{1 \over 4} (\Tr\,g_{(1)})g^{i}_{(1) j}
\Bigg] \nn\\&&  +\CO(e^{-3 \eta}),
\end{eqnarray}
where we have kept the necessary terms up to the relevant orders.

Using the above $\CK$-tensor expressions in conjunction with the
Ricci tensor ones in the previous appendix, one obtains EOM
expression as
\begin{eqnarray}
\CE^{\eta}_{\, \eta} &=& \Big(\sigma -{1 \over {\ell^2}} -{1 \over
{4 m^2}} \Big) + e^{-\eta} \Big(- { \sigma \over 2} +{1 \over {4
m^2}} \Big) \,
\Tr\,g_{(1)} \nonumber \\
&& + e^{-2 \eta} \bigg[ \Big( - {\sigma \over 2} + {1 \over 4 m^2}
\Big) \left(R_{(0)} + 2 \Tr\,g_{(2)} \right) + \Big( { {3\, \sigma}
\over 8} -{1\over {16 m^2}} \Big)\, \Tr\,g^2_{(1)} + \Big({\sigma
\over 8} - {3
\over {16 m^2}} \Big) (\Tr\, g_{(1)})^2 \bigg] \nonumber \\
&& + e^{-3 \eta}  \bigg[ \Big( -{\sigma \over 2} +{1\over {4m^2}}
\Big) \Big( \bar{\nabla}^{i}\bar{\nabla}^{j} g_{(1)\, ij}
-\bar{\nabla}^2\Tr\,g_{(1)} \Big)+\Big( {\sigma \over 4} -{ 1 \over
{4
m^2}} \Big) R_{(0)}\, \Tr\,g_{(1)} \nonumber \\
&&\quad \qquad + \Big(\sigma - {1 \over {2 m^2}} \Big) \Big(
\Tr(g_{(1)} g_{(2)}) - {3 \over 2} \, \Tr\,g_{(3)} \Big) - \Big(
{\sigma \over 4} + {9 \over {32 m^2}} \Big) \, \Tr\,g_{(1)} \,
\Tr\,g^2_{(1)}
\nonumber \\
&& \quad \qquad + \Big( {\sigma \over 2} -{1 \over {2 m^2} }\Big)\,
\Tr\,g_{(1)}\, \Tr\,g_{(2)} - \Big( {\sigma \over 4} - {3 \over {8
m^2}} \Big)\, \Tr\,g^3_{(1)} +{7 \over {32 m^2}} \, (
\Tr\,g_{(1)})^3 \bigg] +\CO(e^{-4 \eta})\,,
\\
%
\CE^{\eta}_{\, j} &=& e^{- \eta} \bigg[ \Big({\sigma \over 2} - {1
\over {4m ^2}} \Big) \Big(\bar{\nabla}_{i}\Tr g_{(1)}
-\bar{\nabla}_{j}g^{j}_{(1)\, i}  \Big) \bigg] + e^{-2 \eta} \bigg[ \Big( \sigma+{1 \over {2 m^2}} \Big) \Big(\bar{\nabla}_{i}\Tr g_{(2)} -\bar{\nabla}_{j}g^{j}_{(2)\, i}   \Big)  +\CO(g^2_{(1)}) \bigg]  \nn \\
&& + \CO(e^{-3 \eta})  \,,
\\
%
\CE^{i}_{\, j} &=& \Big(\sigma -{1 \over {\ell^2}} -{1 \over {4m^2}}
\Big) \,\eta^{i}_{j}  +e^{-\eta} \bigg[ \Big({\sigma \over 2} - {1
\over {4m^2}} \Big) \Big( g^{i}_{(1) j} - (\Tr\,g_{(1)})
\,\eta^{i}_{j}
\Big) \bigg] \nonumber \\
&& +e^{-2 \eta} \bigg[ \Big(\sigma - {1 \over {2m^2}} \Big) \Big\{
{1 \over 8} \Big( (\Tr\,g_{(1)})^2+\Tr\,g^2_{(1)} \Big) \eta^{i}_{j}
- {1 \over 4} \,(\Tr\,g_{(1)}) \,g^{i}_{(1) j} \Big\}  \nonumber \\
&& \qquad ~\quad + \Big(\sigma +{1 \over {2m^2}} \Big)
\Big(R^{i}_{(0) j} - {1 \over 2} \,R_{(0)} \,\eta^{i}_{j}  \Big)
\bigg] +\CO(e^{-3 \eta})  \,,
\end{eqnarray}
where the two dimension identity $ R^{i}_{(0) j} - {1 \over 2}
\,R_{(0)} \,\eta^{i}_{j}=0 $ is used.

These EOMs imply that  $g_{(1)}$ term  should vanish at the generic
point in the parameter space except the special point $\sigma=2m^2
=1 ={2/ {\ell^2}}$ where the last equality is from Eq.(\ref{rel}).
More concretely,  the first term in
$\CE^{\eta}_{~\eta}=\CE^{i}_{~j}=0$  give us the condition $\sigma -
1/\ell^2  - 1/4m^2 =0$, and the second term gives $g_{(1)} =0$ or
$\sigma= 2m^2$. Therefore, one can see that $g_{(1)}=0$ except the
special point $\sigma=2m^2=1$. At a generic parameter point, the
next order term from EOMs leads to
\[
  R_{(0)} + 2 \Tr\,g_{(2)}  =0 \,.
\]

At the special point of $\sigma=2m^2=1 ={2/ {\ell^2}} $,  the $\eta
\eta$-component of EOMs, $\CE^{\eta}_{\, \eta}=0$, gives the
following two conditions
\bea
(\Tr\,g_{(1)})^2 &=&\Tr\,g^2_{(1)}\,,  \nn \\
\Big( R_{(0)} + 2 \Tr\,g_{(2)} -{1 \over 2} (\Tr\,g_{(1)})^2 \Big)
\,\Tr\,g_{(1)}  &=& 0\,, \nn  \eea
where the first condition comes from $e^{-2\eta}$ order and the
second one from  $e^{-3 \eta}$ order using the relation
$(\Tr\,g_{(1)})^3 = \Tr\,g^3_{(1)}$ derived by the first condition
in two dimensions. Note that the $i j$-component of EOMs,
$\CE^{i}_{j}=0$, does not give a non-trivial condition up to $e^{-2
\eta}$ order, and the $\eta i$-component of EOMs,
$\CE^{\eta}_{\,i}=0$, also gives no condition up to $e^{-\eta}$
order.

Even at the special point $\sigma=2m^2=1$,   $g_{(1)} $ may vanish.
In this case,  the $\eta \eta$-component of EOMs gives no condition
at the  $e^{-3 \eta}$ order. But  one condition can be obtained   at
$e^{-4 \eta}$ order.  By computing the fully contracted EOM, this
fact can be easily seen as follows. Since FG expansion of $\CK$-term
is given by
 \bea
  \CK
  &=& -\frac{3}{2} +  \half e^{-2\eta}\bigg[ R_{(0)} + 2\Tr\, g_{(2)} \bigg] \nn \\
 &&  + \frac{1}{2}e^{-4\eta}\bigg[- 4\Tr\, g_{(4)}   +  \Tr\, g^2_{(2)}
-\frac{1}{4}\big(R_{(0)} +2\Tr\, g_{(2)} \big) \big(  R_{(0)}+4\Tr\,g_{(2)} \big) \nn \\
&&\qquad  \qquad + {g^{ik}_{(0)}\Big(\bar{\nabla}_{l}
\Gamma^{l}_{(2)\, ik}-\bar{\nabla}_{i}\Gamma^{~~ l}_{(2)\, lk}
\Big)}
  \bigg] + \CO(e^{-5\eta})\,, \nn \eea
at these special point, fully contracted EOM expression is given by

 \begin{equation}
   \CE^{\mu}_{\, \mu}   =   -{1 \over 8} \,e^{-4 \eta} \,\Big(R_{(0)} +2\Tr\, g_{(2)} \Big)^2  +O(e^{-5 \eta}) \,.
\end{equation}
Therefore when $g_{(1)}=0$, regardless of the value of the parameter
$m^2$ we have the following condition.
  \begin{equation}
R_{(0)} +2\Tr\, g_{(2)} =0\,.
\end{equation}

\newpage

\section{The Absence of `Log' Term at a Generic Point}
In this appendix, we show that there is no `$\log$' terms, which is
proportial to  $\eta$ in our convention, at a generic point in the
parameter space.
Let us begin with the following Fefferman-Graham expansion with
`$\log$' term,
\begin{equation}
\gamma_{ij}  =  e^{2\eta} g^{(0)}_{ij} + e^{\eta}  g^{(1)}_{ij} +
\eta e^{\eta}  b^{(1)}_{ij}  +  g^{(2)}_{ij} + \eta b^{(2)}_{ij} +
e^{-\eta} g^{(3)}_{ij} +   e^{-2\eta}g^{(4)}_{ij} +
\CO(e^{-3\eta})\,,
\end{equation}
The fully contracted EOM is expressed in the form of
\begin{eqnarray}
   \CE^{\mu}_{\, \mu}   &=&  3 \left( \sigma- {1 \over \ell^2} -{1 \over {4 m^2}} \right) +e^{-\eta} \Bigg[ \left(- \sigma +{1 \over {2m^2}} \right) \left(\Tr\,g_{(1)} - \frac{1}{2} \Tr\,b_{(1)} \right)  \Bigg] \nn \\
&&+ \eta e^{-\eta} \left( -\sigma +{1 \over {2m^2}} \right)
\Tr\,b_{(1)} + \CO(e^{-2\eta}) =0\,.
\end{eqnarray}
At a generic point, one can see that  $\Tr\,b_{(1)}=0$. Since the $i
j $ component of EOM is given by
\begin{eqnarray}
\CE^{i}_{\, j} &=& \Big(\sigma -{1 \over {\ell^2}} -{1 \over {4m^2}}
\Big) \,\eta^{i}_{j}  +e^{-\eta} \bigg[ \Big({\sigma \over 2} - {1
\over {4m^2}} \Big) \Big( g^{i}_{(1) j} - (\Tr\,g_{(1)})
\,\eta^{i}_{j}
\Big) \bigg] \nonumber \\
&&+\eta e^{-\eta} \Bigg[ \left( {\sigma \over 2} -{1 \over {4 m^2}}
\right) \left( b^{i}_{(1) j} -  (\Tr\,b_{(1)})\,\eta^{i}_{j} \right)
\Bigg]+\CO(e^{-2\eta})  =0 \,,
\end{eqnarray}
one can insure that $  b^{i}_{(1) j}  =0$  at a generic point.

Using this result,  the second order of the fully contracted EOM can
be written simply as
\begin{eqnarray}
 && e^{-2\eta} \Bigg[ \left( -{\sigma \over 2}+{1 \over {4m^2}} \right)( R_{(0)} + 2\,\Tr\,g_{(2)} + \Tr\,b_{(2)} ) \Bigg] \nn \\
&&+\eta e^{-2\eta} \left(-\sigma + {1 \over {2m^2}} \right)
\Tr\,b_{(2)} +\CO(e^{-3\eta})=0 \,.
\end{eqnarray}
{}From this,  one can also see that $ \Tr\,b_{(2)} =0$ at  a generic
point.
Since the second order of $i j $ component of EOM is given by
\begin{eqnarray}
e^{-2\eta} \Bigg[   \left(\sigma +{1 \over {2 m^2}} \right)
b^{i}_{(2) j} - \sigma(\Tr\,b_{(2)}) \,  \eta^{i}_{j}
\Bigg]+\CO(e^{-3 \eta}) =0 \,,
\end{eqnarray}
one can also verify  that $  b^{i}_{(2) j}  =0$ at a generic point.

\newpage
\section{The Computation of the On-shell Action Value}
Though it is naturally expected that the divergence cancelation
condition from the on-shell action value is identical with the one
from stress tensor, the on-shell action value is presented in this
appendix. Let us put the $AdS$ boundary at the radius $\eta
=\eta_{\infty}$ near the asymptotic infinity.  This renders finite
the on-shell action value, while the divergence of the on-shell
action value appear by sending $\eta_{\infty}\rightarrow\infty$.
Explicitly, the on-shell action value of NMG in Fefferman-Graham
coordinates is given by
\bea S_{NMG}&=& \frac{1}{16\pi G} \int d^3 x \sqrt{-\gamma}\Big[\sigma R + \frac{2}{\ell^2}+ \frac{1}{m^2}\CK\Big]   \nn \\
&=& \frac{1}{16\pi G}\int d^2x \sqrt{-g_{(0)} }  \Big(\sigma +
\frac{1}{2m^2}\Big)\bigg[-2  e^{2\eta_{\infty}} +  \eta_{\infty}
R_{(0)}  + \cdots  \bigg]\,, \eea
where we have kept the potentially divergent part and $\cdots$
denote the finite part. Note that  the $g_{(1)}$ term doesn't appear
in the divergent part. The above divergent part is completely same
with the pure Einstein gravity case. That is to say, for the counter
action $S_{c.t.}$ one needs just the boundary  cosmological constant
term and $(\eta/16\pi G)\int d^2x \sqrt{-\gamma} R[\gamma]$ term
which is omitted in our paper since it is trivial under metric
variation in two dimension.

It is straightforward to obtain the on-shell value  of generalized
GH terms  as
\bea
 S_{GH} &=& \frac{\xi}{16\pi G }\int d^2x \sqrt{-\gamma}\Big[2\sigma K + \hat{f}^{ij}K_{ij} - \hat{f}K\Big] \nn \\
 &=& \frac{\xi}{16\pi G }\int d^2x \sqrt{-g_{(0)}} \bigg[  4e^{2\eta_{\infty}} \Big(\sigma + \frac{1}{2m^2}\Big) +  e^{\eta_{\infty}}\Big(\sigma -\frac{1}{2m^2}\Big)\Tr\, g_{(1)} \bigg]  + \cdots \,.  ~~~~~ \eea
The on-shell value of counter terms is  given by
\bea S_{c.t.} &=&   \frac{1}{8\pi G} \int d^2x \sqrt{-\gamma}~(A +B \hat{f} +C\hat{f}^2 +Df_{ij}f^{ij})   \nn \\
   &=&  \frac{1}{8\pi G} \int d^2x \sqrt{-g_{(0)}}~ \bigg[  e^{2\eta_{\infty}}\Big(A-\frac{2B}{m^2}+\frac{4C}{m^4} + \frac{2D}{m^4}\Big) + e^{\eta_{\infty}} \Big(  \frac{1}{2} A - \frac{2C}{m^4}  - \frac{D}{m^4}\Big)\Tr\, g_{(1)}   \nn \\ && ~~~~~~~~ \qquad \qquad  \qquad  + \cdots \bigg]\,.   \eea
Now, one can ensure that the divergence cancelation condition is
completely identical with stress tensor computation.

\newpage

\end{document}